\documentclass{article}

\usepackage{microtype}
\usepackage{graphicx}
\usepackage{subfigure}
\usepackage{booktabs} 

\usepackage{hyperref}
\usepackage{algorithm}
\usepackage{algpseudocode}
\usepackage{listings}
\usepackage{bm}
\usepackage{bbm}

\usepackage{bbding}
\usepackage{makecell}
\usepackage{multirow}
\usepackage{longtable}
\usepackage{tikz}

\newcommand*\circled[1]{\tikz[baseline=(char.base)]{
            \node[shape=circle,draw,inner sep=1pt] (char) {#1};}}

\usepackage{etoolbox}
\makeatletter
\AfterEndEnvironment{algorithm}{\let\@algcomment\relax}
\AtEndEnvironment{algorithm}{\kern2pt\hrule\relax\vskip3pt\@algcomment}
\let\@algcomment\relax
\newcommand\algcomment[1]{\def\@algcomment{\footnotesize#1}}
\renewcommand\fs@ruled{\def\@fs@cfont{\bfseries}\let\@fs@capt\floatc@ruled
  \def\@fs@pre{\hrule height.8pt depth0pt \kern2pt}%
  \def\@fs@post{}%
  \def\@fs@mid{\kern2pt\hrule\kern2pt}%
  \let\@fs@iftopcapt\iftrue}
\makeatother

\usepackage[accepted]{icml2025}

\newcommand{\ie}{\textit{i}.\textit{e}.}

\usepackage{amsmath}
\usepackage{amssymb}
\usepackage{mathtools}
\usepackage{amsthm}
\usepackage{diagbox}
\usepackage{makecell}
\usepackage{multirow}
\usepackage{multicol}
\usepackage{pifont}
\usepackage{dsfont}

\usepackage[capitalize,noabbrev]{cleveref}

\theoremstyle{plain}

\theoremstyle{definition}

\theoremstyle{remark}
\usepackage{float}

\usepackage[textsize=tiny]{todonotes}

\algblock{ParFor}{EndParFor}
\algnewcommand\algorithmicparfor{\textbf{parfor}}
\algnewcommand\algorithmicpardo{\textbf{do}}
\algnewcommand\algorithmicendparfor{\textbf{end\ parfor}}
\algrenewtext{ParFor}[1]{\algorithmicparfor\ #1\ \algorithmicpardo}
\algrenewtext{EndParFor}{\algorithmicendparfor}

\definecolor{bred}{RGB}{250, 82, 82}
\definecolor{borange}{RGB}{253, 126, 20}
\definecolor{byellow}{RGB}{250, 176, 5}
\definecolor{bgreen}{RGB}{116, 184, 22}
\definecolor{bblue}{RGB}{250, 176, 5}
\definecolor{bindigo}{RGB}{76, 110, 245}
\definecolor{bcyan}{RGB}{59, 201, 219}
\definecolor{bteal}{RGB}{99, 230, 190}

\usepackage{amsmath,amsfonts,bm}

\def\eqref#1{equation~\ref{#1}}

\def\1{\bm{1}}

\DeclareMathAlphabet{\mathsfit}{\encodingdefault}{\sfdefault}{m}{sl}
\SetMathAlphabet{\mathsfit}{bold}{\encodingdefault}{\sfdefault}{bx}{n}

\begin{document}

\twocolumn[
\icmltitle{Hexa-MoE: Efficient and Heterogeneous-aware Training for Mixture-of-Experts}



\icmlsetsymbol{equal}{*}

\begin{icmlauthorlist}
\icmlauthor{Shuqing Luo}{pku}
\icmlauthor{Jie Peng}{ustc}
\icmlauthor{Pingzhi Li}{unc}
\icmlauthor{Hanrui Wang}{ucla}
\icmlauthor{Tianlong Chen}{unc}
\end{icmlauthorlist}

\icmlaffiliation{pku}{Peking University, China}
\icmlaffiliation{ustc}{University of Science and Technology of China, China}
\icmlaffiliation{unc}{University of North Carolina, Chapel Hill, USA}
\icmlaffiliation{ucla}{University of California, Los Angeles, USA}

\icmlcorrespondingauthor{Tianlong Chen}{tianlong@cs.unc.edu}

\icmlkeywords{Machine Learning, ICML}

\vskip 0.3in
]




\begin{abstract}
Mixture-of-experts (MoE) has emerged as a practical approach to scale up parameters for Transformer model to achieve better generalization while keeping computational efficiency. Current MoE models are mainly deployed with expert parallelism on distributed devices, which typically depends on homogeneous devices, and suffers from heavy communication overhead as well as computation redundancy under scaled workload. To tackle these teasers, we propose a \underline{\texttt{H}}eterogeneous-aware \underline{\texttt{EX}}pert \underline{\texttt{A}}llocation framework, \textbf{\texttt{HEXA-MoE}}, with significantly enhanced efficiency. It contains two components: \underline{($1$) \textit{Expert-Specific Operators}}. We replace the typical GeMM or grouped GeMM interfaces with our proposed expert-specific operators, making expert computing to be performed in an in-place manner with almost no redundant FLOPs. \underline{($2$) \textit{Adaptive Data- and Model-Centric}} configurations for different workload scales. We introduce a memory-efficient computation-communication overlapping scheme to tackle the heavy memory consumption in current data-centric libraries, which can accelerate training with heavy workloads. Comprehensive experiments on the Swin-MoE benchmark consistently reveal the effectiveness of \texttt{HEXA-MoE}, \textit{i.e.}, reducing $10\%\sim48\%$ memory consumption and achieving $0.5\sim4.3\times$ speed up compared to current state-of-the-art MoE libraries. We further examine \texttt{HEXA-MoE} on heterogeneous devices, and promising results show that employing optimal parallel configuration can better utilize global computation resources, and substantially minimize overall latency. Codes are available at \href{https://github.com/UNITES-Lab/HEXA-MoE}{\underline{here}}.
\end{abstract}

\section{Introduction}

\begin{figure}[t]
    \centering
    \includegraphics[width=0.999\linewidth]{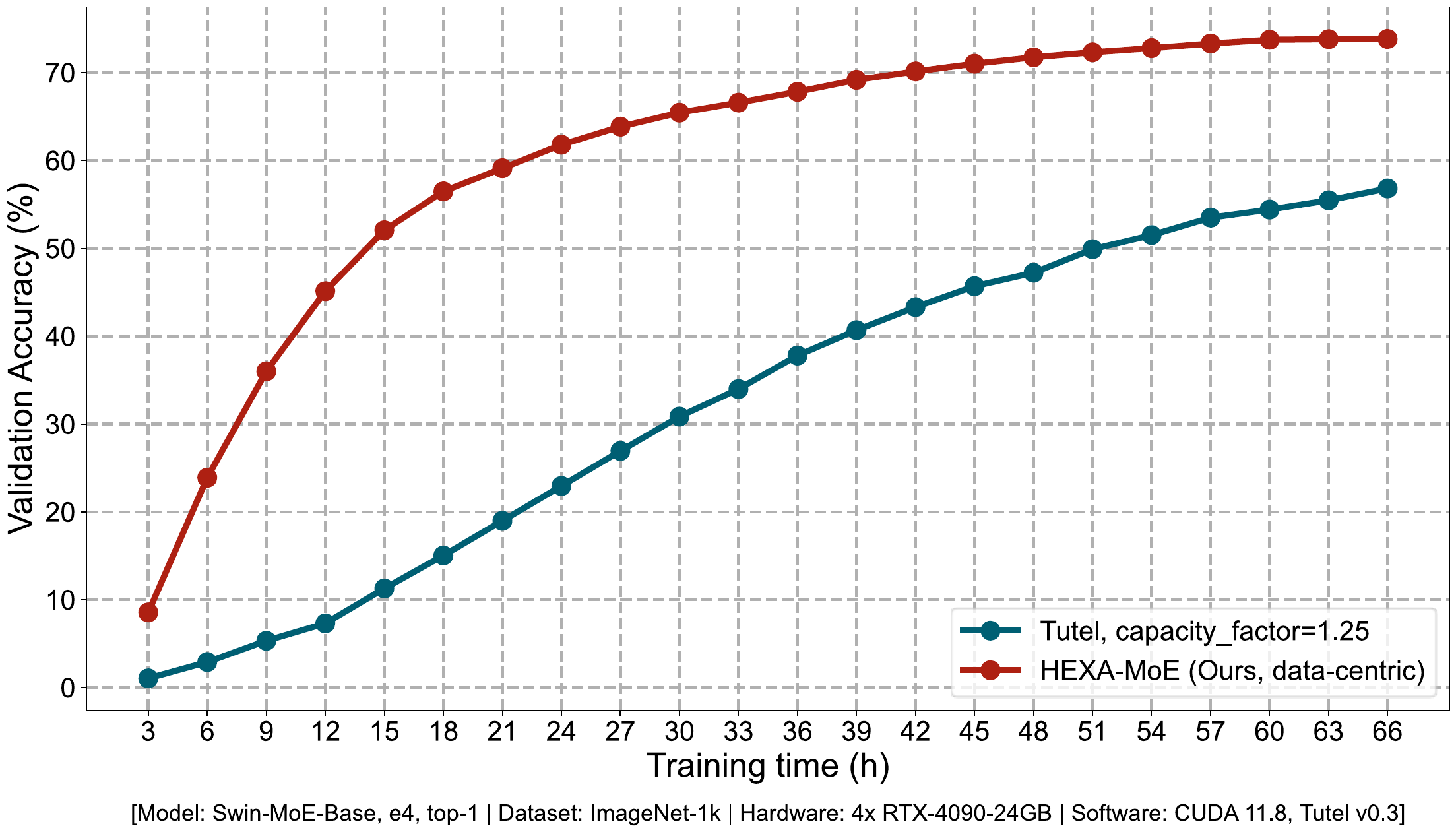}
    \vspace{-0.8cm}
    \caption{\textbf{Convergence Analysis.} \texttt{HEXA-MoE} can significantly surpass Tutel on MoE training due to the specialized designs.}
    \vspace{-0.4cm}
    \label{fig:val-acc}
\end{figure}

\vspace{0.6cm}
\printAffiliationsAndNotice{}

Transformers have become the \textit{de facto} architecture for a vast range of machine learning tasks including natural language process~\cite{vaswani2017attention, raffel2020exploring}, computer vision~\cite{liu2021swin, he2022masked} and multi-modal learning~\cite{li2022blip, liu2024visual}. To scale up model parameters for stronger learning capacity and better generalization following the empirical scaling law~\cite{kaplan2020scaling}, Mixture-of-experts (MoE)~\cite{jiang2024mixtral} has been demonstrated as a practical and computation-efficient approach. For a single Transformer layer, MoE expands the feed-forward network (FFN) to a group of experts with the same architecture and different parameters, activated selectively during computing. This dynamic nature poses peculiar challenges to system design. Current MoE frameworks usually present some unique attributes that differentiate them from dense models:

\begin{enumerate}
    \vspace{-0.5cm}
    \item [\ding{182}] Distributed computing is required due to its sheer size, and \underline{expert parallelism}~\cite{lepikhingshard} is the most commonly used technique, which distributes the experts in one layer on different devices. \vspace{-0.3cm}
    \item [\ding{183}] The dynamic workload for each expert makes it necessary to employ additional \underline{dispatch and combine} operations in expert parallel to utilize the general matrix multiplication (GeMM) or grouped GeMM interface.
    \vspace{-0.3cm}
    \item [\ding{184}] Dispatch and combine rely on \underline{synchronous all-to-all} communication to assign tokens to distributed experts.
\end{enumerate}
\vspace{-0.5cm}

These peculiar designs lead to apparent inefficiency for MoE training. \underline{On the one hand}, current MoE libraries are either static like Tutel~\cite{hwang2023tutel}, or dynamic like MegaBlocks~\cite{gale2023megablocks}. The former depends on an additional hyper-parameter named expert capacity to calibrate the workload for different experts, which suffers from redundant memory allocation or access, while the latter tends to raise out of memory (OOM) error during runtime due to its uncertainty. Meanwhile, the synchronous all-to-all communication can occupy over $40\%$ runtime with scaled model and devices~\cite{hwang2023tutel}. \underline{On the other hand}, expert parallelism is mainly deployed on homogeneous devices. However, a homogeneous cluster with cutting-edge devices is much more expensive than heterogeneous ones due to the fast iteration of modern GPU and the decrease on prices for outdated ones. Adapting expert parallelism to \textit{heterogeneous} devices would rely on rescheduling expert placement to utilize global computing resources sufficiently. However, the inherent dynamic property of MoE makes it difficult to implement, as the workload of each expert changes dynamically in each step~\cite{he2022fastermoe, li2023accelerating}. Since heterogeneous devices are cheaper and easier to access, if MoE training can be refactored to be heterogeneous-aware, it would advance wider deployment.

We propose {\texttt{HEXA-MoE}}, a completely-static MoE training framework with minimized memory consumption and heterogeneous-awareness. We first find that if implementing MoE layer with GeMM, the unnecessary components such as token padding or discarding are unavoidable, while grouped GeMM~\cite{gale2023megablocks} would lead to dynamic runtime behavior and uncertainty. To tackle it, we investigate the forward and backward propagation for the MoE layer thoroughly, and propose to replace (grouped) GeMM with \textit{expert-specific} operators. Specifically, the basic MoE operations are essentially build upon $3$ operators, \ie, expert-specific matrix multiplication (\textit{ESMM}), summation (\textit{ESS}) and transposed matrix multiplication (\textit{ESTMM}). The forward propagation only requires \textit{ESMM}, while backward requires all. We implement them with specialized GPU kernels, and distribute the MoE layer with tensor parallelism~\cite{narayanan2021efficient} as an alternative to expert parallelism, which is easy to be heterogeneous-aware.

\texttt{HEXA-MoE} also considers different workload scales for MoE training, and is adapted to both data- and model-centric configurations. Their difference lies in the content of communication. For model-centric, local mini-batches are synchronized among devices and model parameters are kept still, while for data-centric, local parameters are gathered for each device, while local mini-batches are kept still~\cite{liu2023janus}. Data-centric setting outperforms model-centric with scaled workloads. However, previous library preserves all the gathered parameter shards on each device for backward pass, leading to huge memory consumption. To tackle this teaser, we design a novel memory-efficient communication-communication overlapping scheme utilizing a pipeline-shared cache on each device, which is updated dynamically for both forward and backward pass. Both settings are completely static in runtime, and the workload of each device can be easily determined from batch size or tensor parallelism configuration~(sub-dimension of FFN intermediate size), making it easy to be adapted to heterogeneous devices via specialized expert allocation. 

Our contributions can be summarized as follows:
\vspace{-0.4cm}
\begin{enumerate}
    \item [$\star$] \textbf{A completely static MoE training framework with minimized memory footprint}. To our best knowledge, \texttt{HEXA-MoE} is the first MoE library with completely static and minimal memory footprint, enabling in-place computing with specialized kernel, and employing tensor parallelism for distributed training, delivering exact result with faster speed, as shown in Figure~\ref{fig:val-acc}.
    \vspace{-0.3cm}
    \item [$\star$] \textbf{Advanced efficiency}. \texttt{HEXA-MoE} is built upon our proposed expert-specific operators, which tackles scaled workload with data-centric strategy via memory-efficient communication-computation overlapping. Experiments on Swin-MoE show that \texttt{HEXA-MoE} can reduce $10\%\sim48\%$ memory usage and achieve $0.5\sim4.3\times$ speed up compared to SOTA MoE libraries.
    \vspace{-0.3cm}
    \item [$\star$] \textbf{Heterogeneous-awareness}. \texttt{HEXA-MoE} transforms conventional expert parallelism into data or tensor parallelism, making it heterogeneous-aware. Experiments show that \texttt{HEXA-MoE} can be effectively adapted to heterogeneous devices and substantially minimize overall latency via employing optimal parallel configuration.
\end{enumerate}
\vspace{-0.3cm}

\vspace{-0.3cm}
\section{Related Works} 

\begin{figure*}
    \centering
    \includegraphics[width=2.06\columnwidth]{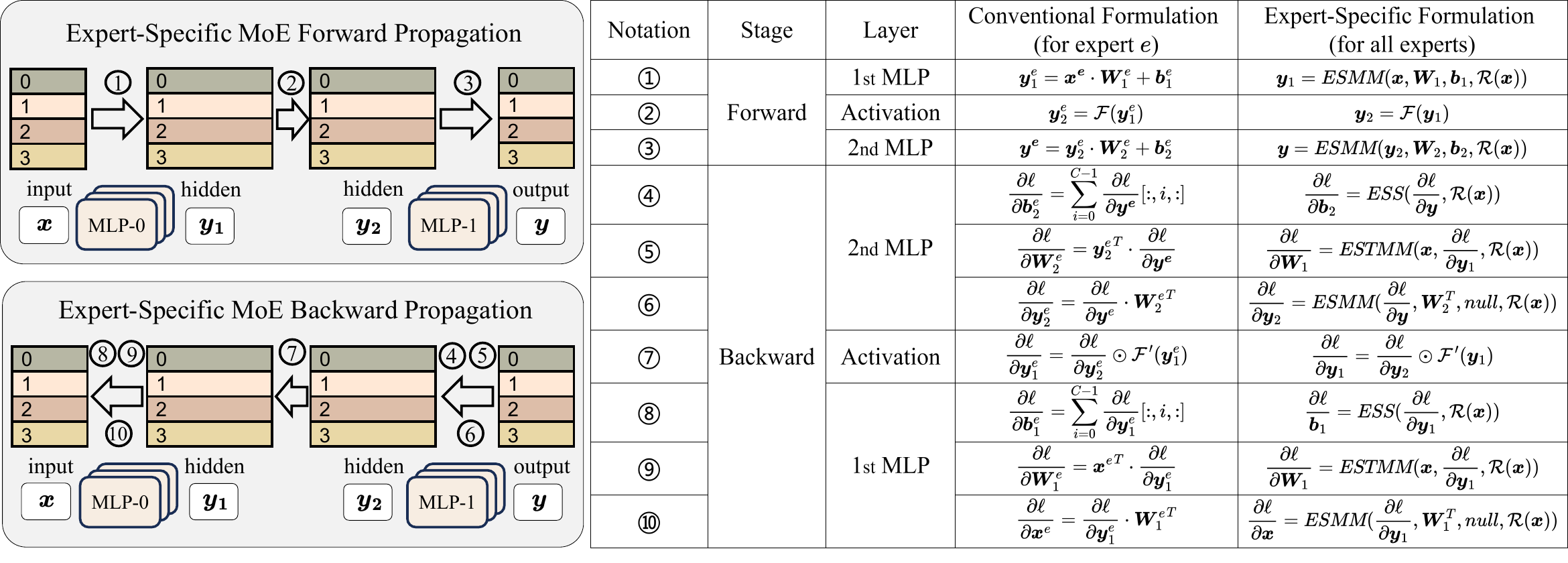}
    \vspace{-0.5cm}
    \caption{\textbf{Comparison between conventional and expert-specific formulation for MoE computing.} We take top-1 routing for illustration and present the corresponding relation of each formula in the MoE forward and backward propagation.}
    \label{fig:formula}
    \vspace{-0.5cm}
\end{figure*}

\vspace{-0.2cm}
\paragraph{Scaling Transformers with MoE.} Transformer models can be scaled up via the Mixture-of-Experts~(MoE) for better learning capacity and generalization while enjoying a sub-linear increase in computation overhead due to its sparsity. It has been proven in many research fields such as natural language processing~\cite{fedus2022switch, jiang2024mixtral, du2022glam}, machine vision~\cite{riquelme2021scaling, fan2022m3vit} and multi-modal learning~\cite{bao2022vlmo}. 
\vspace{-0.4cm}
\paragraph{Open-Source MoE Training Libraries.} FastMoE~\cite{he2021fastmoe} pioneered the first PyTorch open-source implementation for distributed MoE. Based on it, FasterMoE~\cite{he2022fastermoe} proposes dynamic shadowing and smart scheduling to tackle load imbalance and improve parallelism. Tutel~\cite{hwang2023tutel} implements switchable parallelism and dynamic pipelining, improving adaptability and scalability. MegaBlocks~\cite{gale2023megablocks} proposes block-sparse operations and corresponding GPU kernels to mitigate dynamic routing issues in MoEs.
\vspace{-0.4cm}
\paragraph{Improving efficiency for MoE computing.} MoE computing faces significant challenges in communication and memory. Recent research has explored alleviating them from various perspectives: Lina~\cite{li2023accelerating} proposes to dynamically schedule resources to reduce latency and tackle the all-to-all bottleneck. Janus~\cite{liu2023janus} proposes a data-centric paradigm that replaces traditional all-to-all communication with asynchronous expert fetching, enabling overlapped communication and computation for enhanced parallelism. ScheMoE~\cite{shi2024schemoe} proposes a generic scheduling framework for optimal communication and computation scheduling during MoE training. MPMoE~\cite{zhang2024mpmoe} accelerates MoE training through adaptive and memory-efficient pipeline parallelism. SmartMoE~\cite{zhai2023smartmoe} proposes an efficient searching algorithm to identify optimization opportunities within an expanded hybrid parallelism space, tailored for data-sensitive MoE models. PipeMoE~\cite{shi2023pipemoe} adaptively pipelines communications and computations in MoE to mask communication latency. It also provides an optimal strategy to determine pipeline degree to minimize overall iteration time. 
\vspace{-0.6cm}
\paragraph{Principles of GPU Acceleration} Modern GPUs provide massive threads for parallel execution, grouped into \textit{thread-block}s, and executed on streaming multiprocessors~(SMs). GPUs have a memory hierarchy, outlined as large but slow-accessed high bandwidth memory (HBM), and small but faster-accessed shared memory (SRAM). Matrix multiplication is optimized on GPU with \textit{tiling}, \textit{i.e.}, partitioning the output matrix into small $2$D blocks, where each block is computed by a \textit{thread-block} in parallel. The size of individual blocks can be adjusted to improve runtime performance.


\vspace{-0.4cm}
\section{Method}
\vspace{-0.2cm}
We first formulate the forward and backward propagation for a single MoE layer with GeMM and expert-specific operators respectively~(Sec.~\ref{sec:formulation}), which deliver exact results in different manner. After that, we expound the details for our specialized kernel (Sec.~\ref{sec:cuda}), where a novel expert-specific fused operator is introduced for parallel MoE backward pass. These empower \texttt{HEXA-MoE} with high computational efficiency. Next, we consider different workload scales, and adapt \texttt{HEXA-MoE} efficiently to both \textit{data-} and \textit{model-}centric settings (Sec.~\ref{sec:scale}). Finally, we adapt \texttt{HEXA-MoE} to heterogeneous devices, and provide an expert allocation approach to minimize the average latency so as to utilize better heterogeneous computing capacities (Sec.~\ref{sec:hete}).

\vspace{-0.2cm}
\subsection{MoE Computing Formulation}\label{sec:formulation}
\vspace{-0.2cm}
\paragraph{MoE Computing with GeMM.} To employ GeMM for expert computing, we need to calibrate the amount of dispatched tokens for each expert via token padding or discarding. Taking top-1 routing as an example, we formulate the forward and backward propagation in Figure~\ref{fig:formula}, denoting $\ell$ as loss value and $\boldsymbol{x}^e$ as the $N_{\textrm{e}}$ tokens dispatched to expert $e$. In backward propagation, the gradients of the output have been provided by the auto-differentiation program.

Based on dispatch and combine, all variables can be derived via basic matrix operations such as summation and multiplication. Although we can utilize the high-performance GeMM interface, these operations are memory inefficient, since the workload of each expert varies dynamically in each step, and token padding or discarding has to be employed to construct new mini-batches for local experts. It can be overcome by our expert-specific design.
\vspace{-0.4cm}
\paragraph{MoE Computing with Expert-Specific Operators.} Based on the above formulations, we propose to refactor the MoE workflow in an in-place manner to address the teaser of inefficiency and dynamic. Specifically, we find that the forward and backward propagation of a single MoE layer can be reformulated with $3$ basic specialized operators from an expert-specific perspective, namely \textit{expert-specific} operators. We take top-$1$ routing for illustration, while formulations for top-$k$ routing are provided in Appendix~\ref{appendix:algorithm}.

\begin{figure*}[t]
\centering
\label{Fig:ES-Ops}
\begin{minipage}{1.0\linewidth}
    \centering
    \subfigure[Re-Index Vector Construction]{
        \includegraphics[width=0.32\linewidth,height=1.8in]{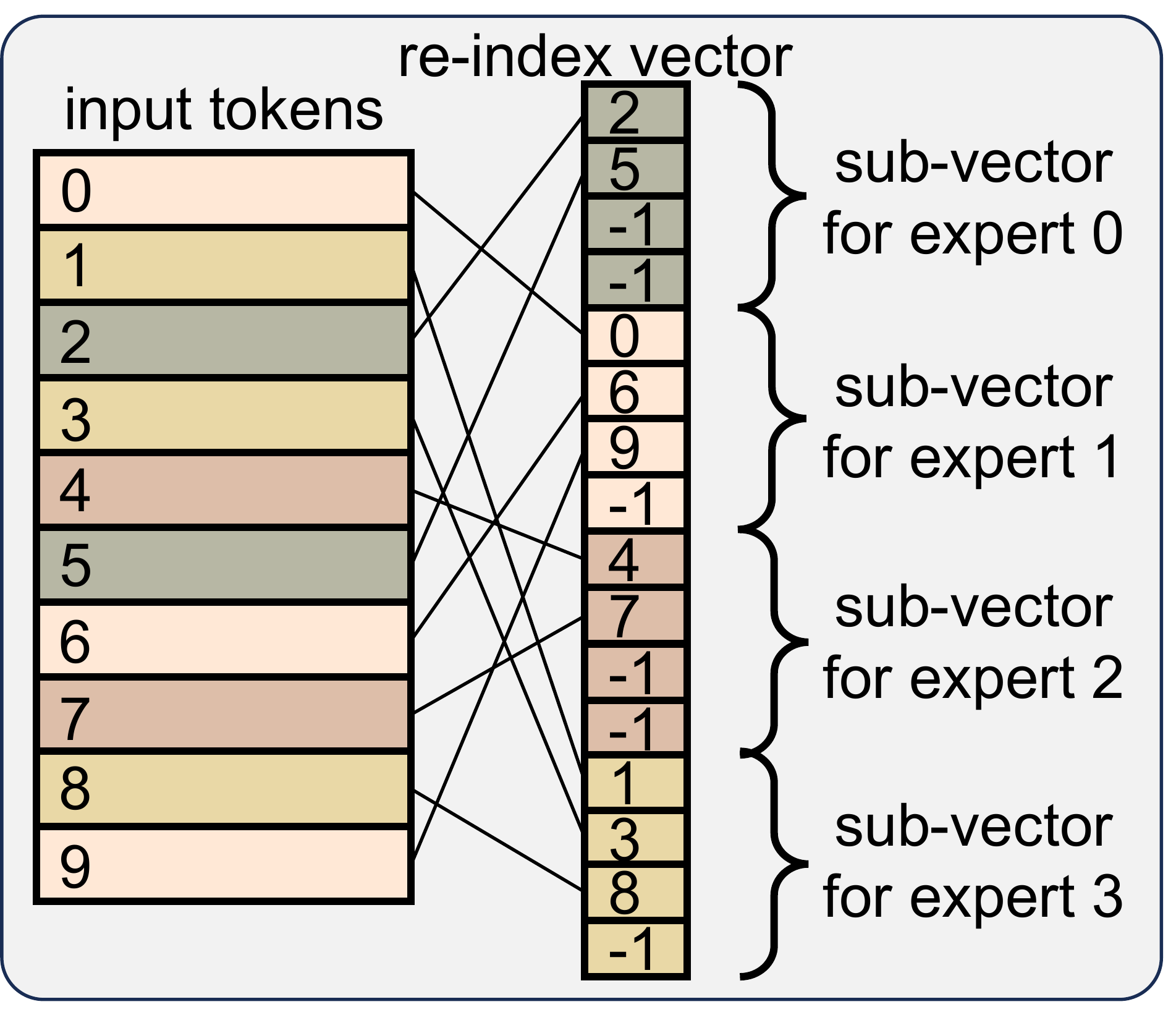}
        \label{fig:re-index}
    }\noindent
    \subfigure[Expert-Specific Matrix Multiplication]{
        \includegraphics[width=0.56\linewidth,height=1.8in]{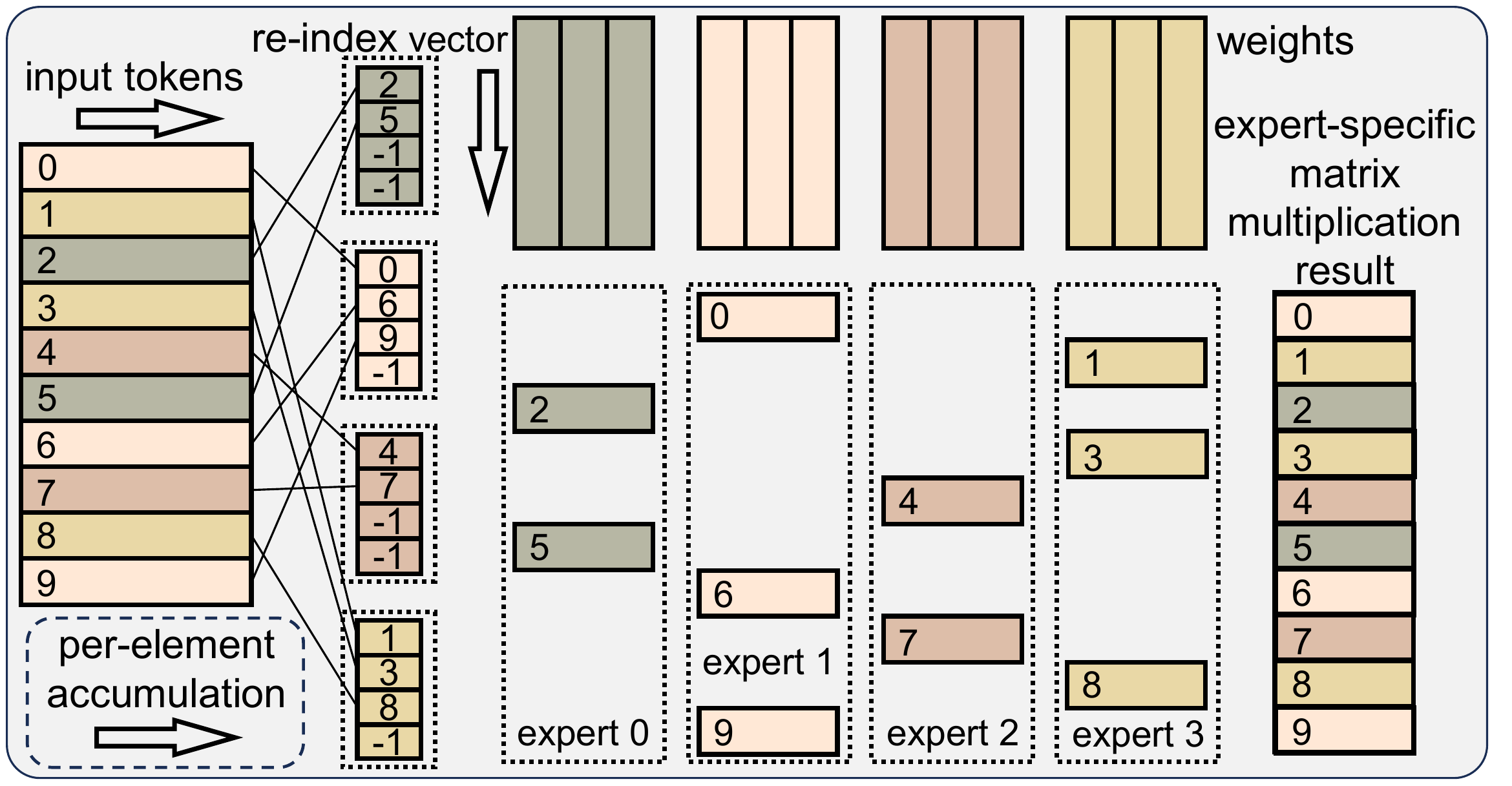}
        \label{fig:esmm}
    }
\end{minipage}\vspace{-0.3cm}
\begin{minipage}{1.0\linewidth}
    \centering
    \subfigure[Expert-Specific Summation]{
        \includegraphics[width=0.44\linewidth,height=1.8in]{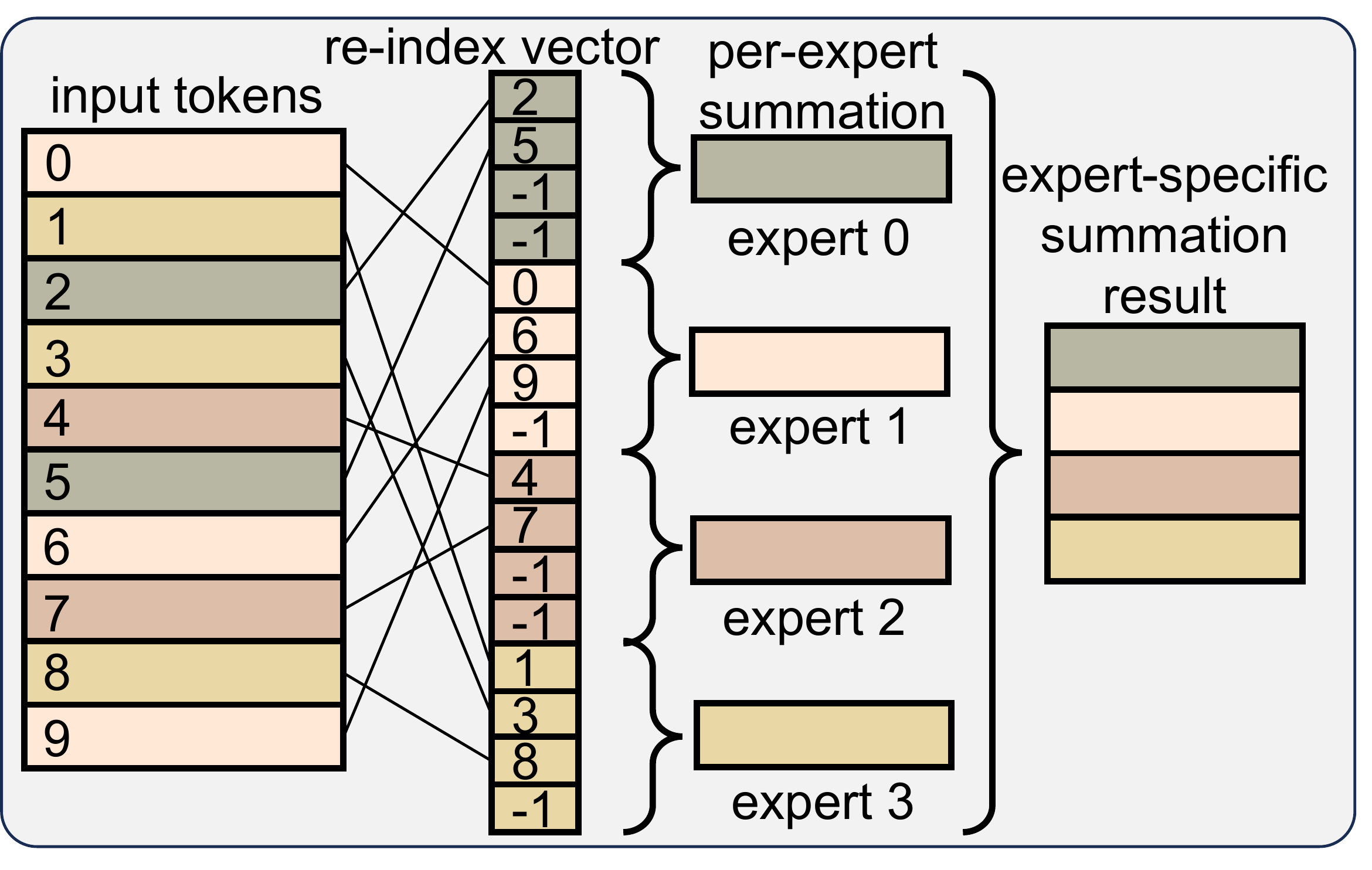}
        \label{fig:ess}
    }\noindent
    \subfigure[Expert-Specific Transposed Matrix Multiplication]{
        \includegraphics[width=0.54\linewidth,height=1.8in]{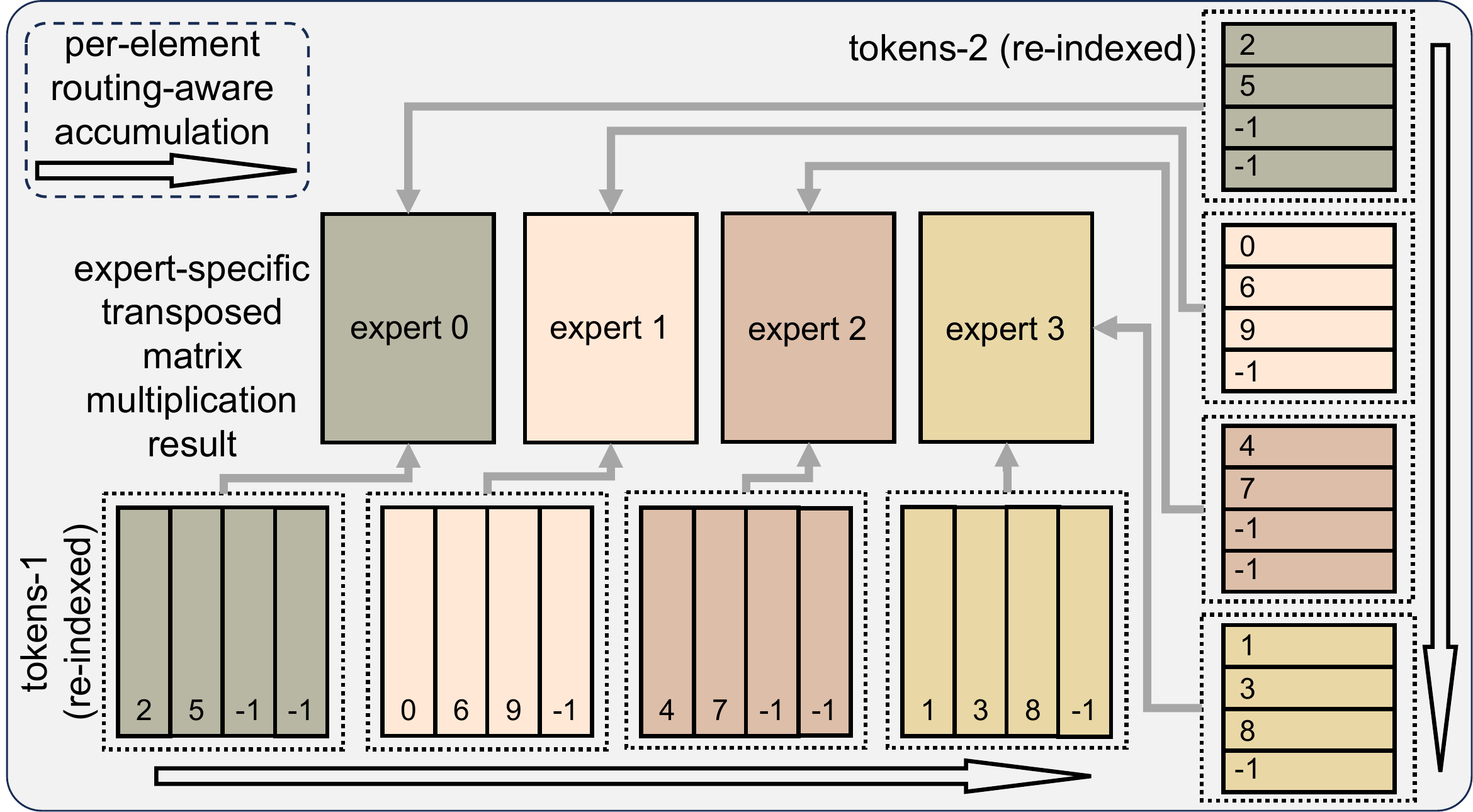}
        \label{fig:estmm}
    }
\end{minipage}
\vspace{-0.7cm}
\caption{\textbf{Illustration of the proposed operators}. We take $10$ tokens, $4$ global experts, and tiling size $4$ as an example. For \textit{ESTMM}, the $2$ input batches are in a re-indexed format, while for others both the raw batch and re-index vector are provided.}
\vspace{-0.6cm}
\end{figure*}

\vspace{-0.5cm}
\begin{enumerate}
    \item [\ding{182}] \textbf{Expert-specific matrix multiplication ($\textit{ESMM}$)}: Given input $\boldsymbol{x}$ with $N$ tokens, routing choice $\mathcal{R}(\boldsymbol{x})$, weight $\boldsymbol{W}$ and bias $\boldsymbol{b}$, the output $\boldsymbol{y}=\textit{ESMM}(\boldsymbol{x}, \boldsymbol{W}, \boldsymbol{b}, \mathcal{R}(\boldsymbol{x}))$, where $\boldsymbol{y}_i$ is derived from $\boldsymbol{x}_i$, $\boldsymbol{W}_{\mathcal{R}(\boldsymbol{x}_i)}$ and $\boldsymbol{b}_{\mathcal{R}(\boldsymbol{x}_i)}$.
    \vspace{-0.3cm}
    \item [\ding{183}] \textbf{Expert-specific summation ($\textit{ESS}$)}: Given input $\boldsymbol{x}$ with $N$ tokens and routing choice $\mathcal{R}(\boldsymbol{x})$, the output $\boldsymbol{y}=\textit{ESS}(\boldsymbol{x}, \mathcal{R}(\boldsymbol{x}))$, where tokens routed to expert $e$ are added up and recorded in $\boldsymbol{y}[e]$.
    \vspace{-0.3cm}
    \item [\ding{184}] \textbf{Expert-specific transposed matrix multiplication ($\textit{ESTMM}$)}: the inputs include $\boldsymbol{x}_1$ and $\boldsymbol{x}_2$, both with $N$ tokens, and the routing choice $\mathcal{R}(\boldsymbol{x})$. Both $\boldsymbol{x}_1$ and $\boldsymbol{x}_2$ are the \textit{ESMM} result of $\boldsymbol{x}$, thus sharing the same routing choice. The output $\boldsymbol{y}=\textit{ESTMM}(\boldsymbol{x}_1, \boldsymbol{x}_2, \mathcal{R}(\boldsymbol{x}))$. For the $i$-th channel of $\boldsymbol{x}_1$ and $j$-th channel of $\boldsymbol{x}_2$, we first prepare a zero vector $\boldsymbol{c}$ with $E$ elements, and accumulate $\boldsymbol{x}_1[m, i]\cdot\boldsymbol{x}_2[m, j]$ to $\boldsymbol{c}[\mathcal{R}(\boldsymbol{x}_m)]$ for all $0\leq m < N$. After that we assign $\boldsymbol{c}[k]$ to $\boldsymbol{y}[k, i, j]$ for all $0\leq k < E$.
\end{enumerate}
\vspace{-0.6cm}
Based on these \textit{expert-specific} operators, we can now implement MoE computing in a novel in-place manner, as formulated in Figure~\ref{fig:formula}. We compare each stage between our method and conventional formulation, and the output of each token is ensured to be consistent for both cases. In forward pass, $\textit{ESMM}$ serves as an alternative to GeMM, while in backward pass, the gradient of $\boldsymbol{y}$ is also provided by the auto-differentiation program. \textit{ESS} performs as an alternative to tensor summation, while \textit{ESMM} and \textit{ESTMM} are designed as $2$ specialized cases for matrix multiplication from the expert-specific perspective. The expert-specific design enables \texttt{HEXA-MoE} to be a completely static framework at runtime, introducing almost no redundant FLOPs.

\vspace{-0.3cm}
\subsection{Optimized CUDA Implementations}\label{sec:cuda}


\vspace{-0.2cm}
\paragraph{Re-Index based Expert-Specific Operators.} To implement tiled matrix multiplication for the expert-specific operators, the HBM I/O should be re-directed, since naively implement them cannot fully utilize GPU locality as well as tensor cores, which restricts runtime performance. In \texttt{HEXA-MoE}, we introduce re-index vector as I/O guidance, illustrated in Figure~\ref{fig:re-index}. Specifically, we re-organize the order of token indices based on the routing choice, \ie, gathering the token indices routed to the same expert into a sub-vector, and padding it with -1 to make it divisible by the tiling size. Algorithm details are provided in Appendix~\ref{appendix:algorithm}.

\underline{For \textit{ESMM}}, we illustrate it in Figure~\ref{fig:esmm}, where a \textit{thread-block} first loads a sub-vector, followed by input tokens based on the vector values, as well as the corresponding expert parameters. Since the loaded tokens are routed to the same expert, we only need to load weights for one expert. We then accumulate the dot product results along the dimension of the input feature, and write the result back to HBM guided by the sub-vector. \underline{For \textit{ESS}}, we illustrate it in Figure~\ref{fig:ess}, where each \textit{thread-block} is assigned with certain channels of one expert, with tokens specialized by the sub-vectors for that expert. After accumulating all the assigned tokens, it writes the result back to HBM. \underline{For \textit{ESTMM}}, we illustrate it in Figure~\ref{fig:estmm}, where input batches are presented in a re-indexed manner. The two inputs here are the $\textit{ESMM}$ results of the same tensor, sharing the same re-index vector. Each thread block loads certain channels of both inputs for tokens routed to the same expert. The cumulative dot production result is write back to the corresponding area on HBM after computing. More algorithm details of this section are provided in Appendix~\ref{appendix:algorithm}.
\vspace{-0.6cm}
\paragraph{Memory Optimization for Top-$k$ Routing.} We find that when extending the routing policy from top-$1$ to top-$k$, the memory footprint would increase significantly. To tackle it, we only enlarge the memory allocation for the intermediate tokens to $k$ times. The input and output tensors (and their gradients) are the cumulation of $k$ \textit{ESMM} results, which can either be implemented in serial, or employ the atomic add interface\footnote{Some data types may not be supported for \texttt{atomic\_add} on NVIDIA GPUs. We initialize the output tensor with float32, and transform it to the target type after computing.}, which lead to similar runtime. 

\begin{figure}[t]
    \centering
    \begin{minipage}{1.0\linewidth}
    \centering
    \subfigure[Training pipeline under data-centric configuration.]{
        \label{fig:pipeline}
        \includegraphics[width=0.98\linewidth,height=1.1in]{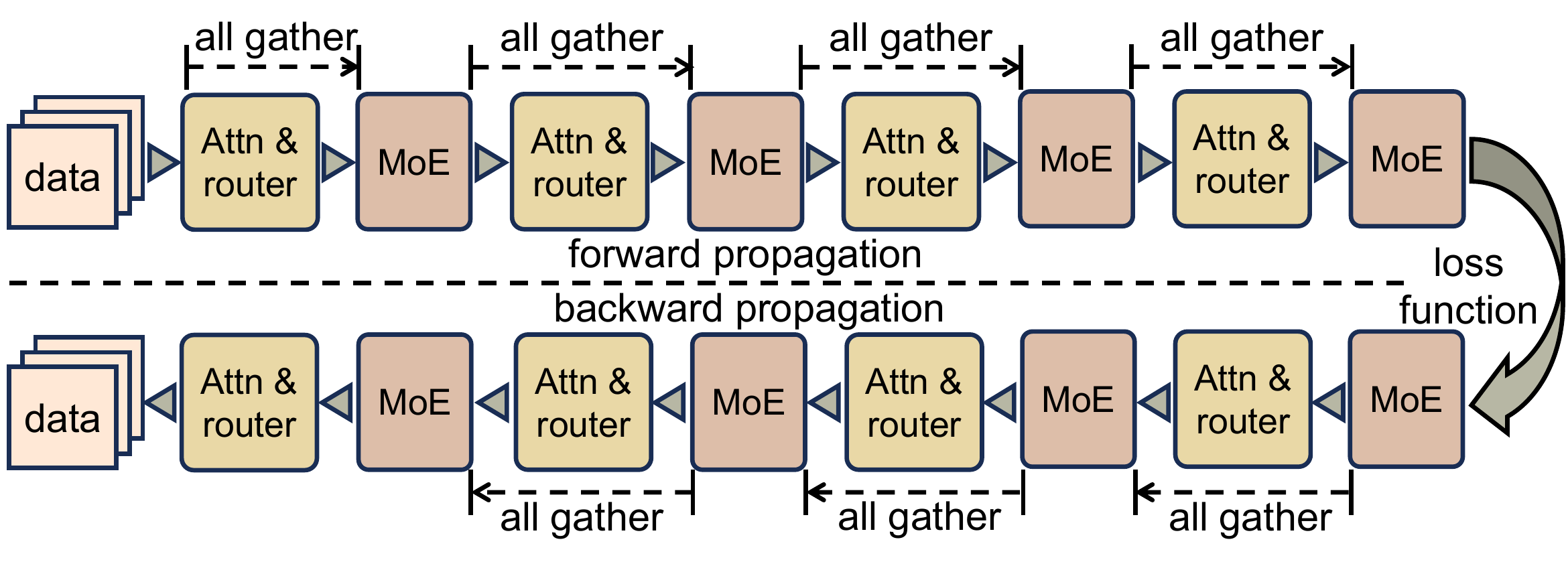}
    }
    \end{minipage}\vspace{-0.2cm}
    \begin{minipage}{1.0\linewidth}
    \centering
    \subfigure[Pipeline-shared cache before and after all gather.]{
        \label{fig:cache}
        \includegraphics[width=0.98\linewidth,height=1.5in]{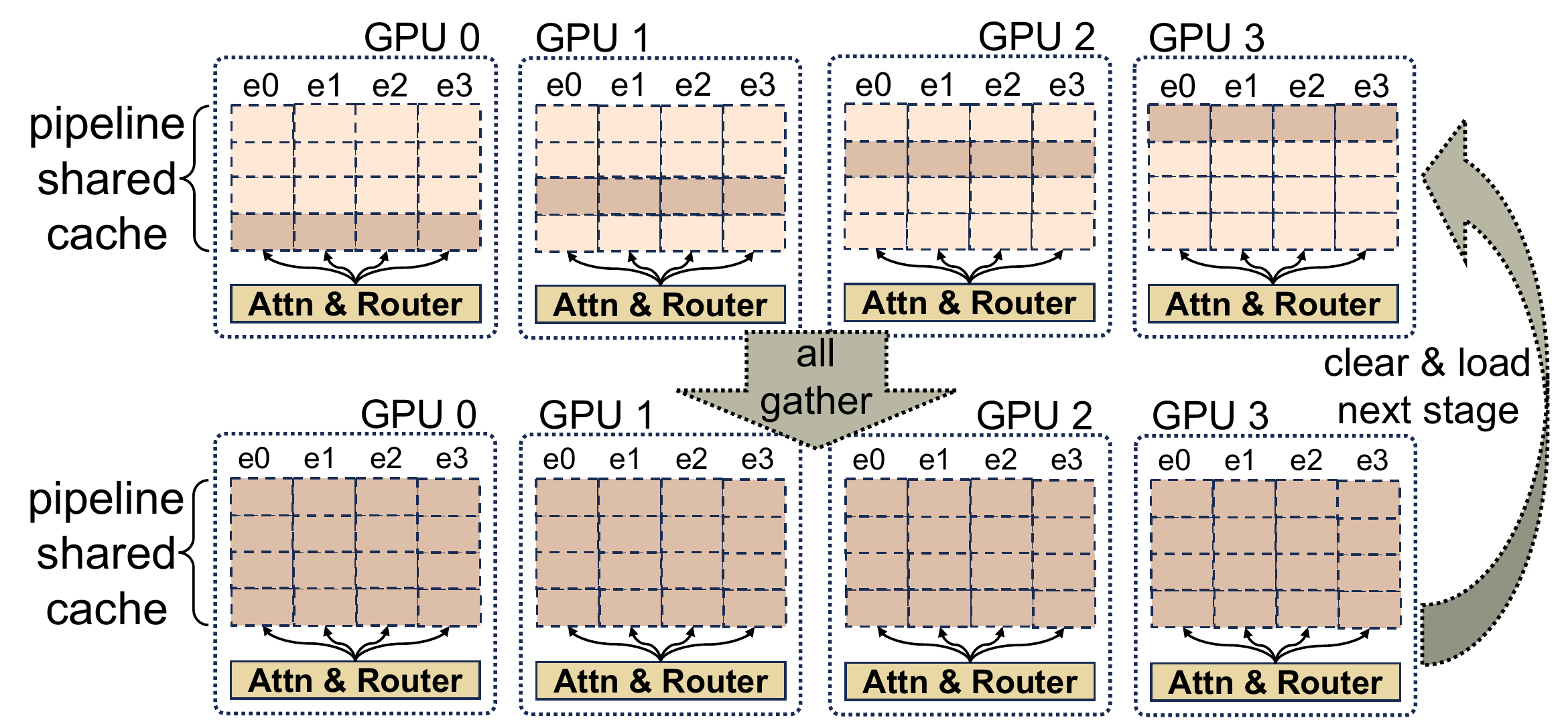}
    }
    \end{minipage}
    \label{fig:pipeline&cache}
    \vspace{-0.6cm}
    \caption{\textbf{Visualization of the training pipeline and shared cache in data-centric setting.} Each device copies the kept parameter shard to the cache before all gather communication, and after that it can access the whole parameters of an MoE layer.}
    \vspace{-0.6cm}
\end{figure}


\vspace{-0.6cm}
\paragraph{Fused Kernel in Backward Propagation.} For a single layer MLP $\boldsymbol{y}=\boldsymbol{x}\cdot\boldsymbol{W}+\boldsymbol{b}$, the gradients for $\boldsymbol{x}$, $\boldsymbol{W}$ and $\boldsymbol{b}$ can be computed in parallel. Similarly, we can integrate \textit{ESS}, \textit{ESTMM}, and \textit{ESMM} together into one kernel, namely expert-specific fused kernel (\textit{ESFK}) for enhanced parallelism. However, directly integrating them is difficult, since the shape of \textit{thread-block}s and the implementation details for each operator vary a lot. To this end, we set the shape of \textit{thread-block} for each operator to be the same, and expand one dimension for \textit{ESMM} and \textit{ESS} so that the thread grids are all $3$-dimensional. Taking a single FFN layer under top-$1$ routing as an example, where we present the shape of allocated thread block and thread grid for both forward and backward propagation in Table~\ref{tab:threads} in Appendix~\ref{appendix:details}. By shape transposing or dim expanding for individual \textit{thread-block}s, we can align the 1st and 2nd dim while aggregate the 3rd dim for an integrated \textit{thread-grid}. In this way, the backward pass for an MoE layer can be implemented with only $2$ fused kernels and an element-wise dot production.

\vspace{-0.3cm}
\subsection{Parallel Strategy for Different Workload Scale}\label{sec:scale}
\vspace{-0.1cm}
\paragraph{Data Parallelism for Data-Centric Setting.} Data-centric is a practical and efficient approach for training deep neural network models with heavy workloads. \texttt{HEXA-MoE} distributes MoE parameters to different devices based on the partition of FFN intermediate size, namely tensor parallelism. In data-centric setting, each device gathers the whole MoE parameters of one layer from other devices, and computes locally, similar to data parallelism. Although data-centric MoE training has been explored by Janus~\cite{liu2023janus}, the teaser of memory efficiency has not been fully tackled. Janus pre-fetchs the required MoE parameters for each layer progressively in forward pass, and preserves all of them locally for backward pass so that no communication would be required in backward. However, the sheer size of MoE parameters can lead to a significant increase in memory usage, making it inefficient for deployment. 

\vspace{-0.08cm}
To tackle this issue, we propose to allocate an additional region on HBM of each GPU to dynamically cache the gathered MoE shards for each pipeline stage, named as pipeline-shared cache. Since each device keeps a subset of the FFN intermediate size for all experts in each layer, we employ all gather communication among devices before computing for an MoE layer, preparing for the required full parameters in both forward and backward pass, as illustrated in Figure~\ref{fig:pipeline} and~\ref{fig:cache}. For better parallelism, all gather communication can be overlapped with other operations such as attention and router, as shown in Figure~\ref{fig:pipeline}. In this way, each device would not have to preserve the full MoE parameters for backward pass, while communication overhead can also be overlapped, therefore both memory efficiency and computing efficiency can be achieved.
\vspace*{-0.4cm}
\paragraph{Tensor Parallelism for Model-Centric Setting.} For small scale of workload less than model parameters, model-centric configuration turns out to be more efficient than data-centric. To distribute MoE parameters on multiple devices, we modify the classical tensor parallelism with our proposed expert-specific operators. All gather communication is employed to synchronize local data batches before and after each MoE layer. We also distribute each MoE layer among different devices along the FFN intermediate size for each expert, and during MoE computing, each device computes the all gathered data batches with only the local MoE parameter chunk using \textit{ESMM}, after that the output tokens are all reduced with sum operation in forward pass. During backward propagation, all gather and all reduce communications are interchanged, while \textit{ESMM} are replaced by \textit{ESMM}, \textit{ESS}, and \textit{ESTMM} to get the gradients for input tokens, bias, and weights, respectively, which can also be replaced by the fused operator \textit{ESFK}. 

\begin{figure*}[t]
    \centering
    \begin{minipage}{1.0\linewidth}
    \centering
    \subfigure{
        \includegraphics[width=0.98\linewidth,height=1.3in]{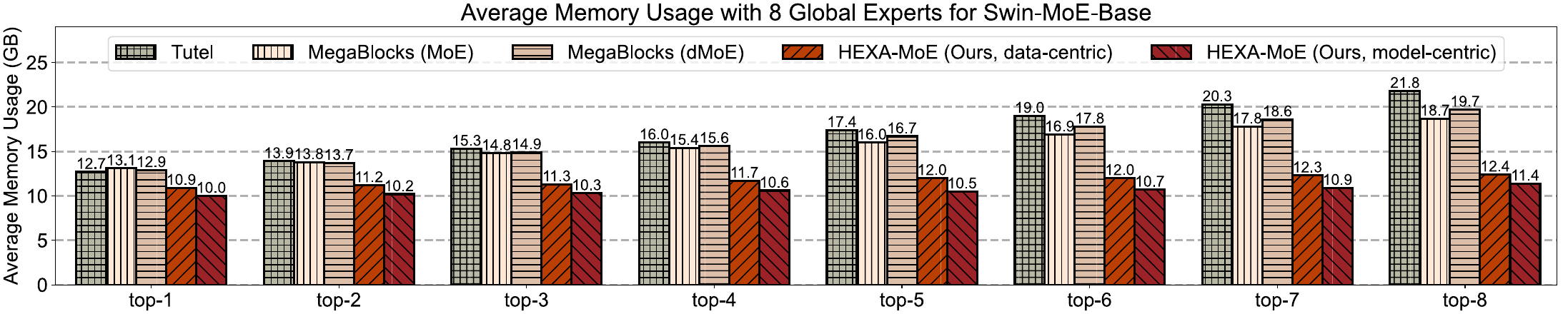}
    }
    \end{minipage}\vspace{-0.2cm}
    \begin{minipage}{1.0\linewidth}
    \centering
    \subfigure{
        \includegraphics[width=0.98\linewidth,height=1.3in]{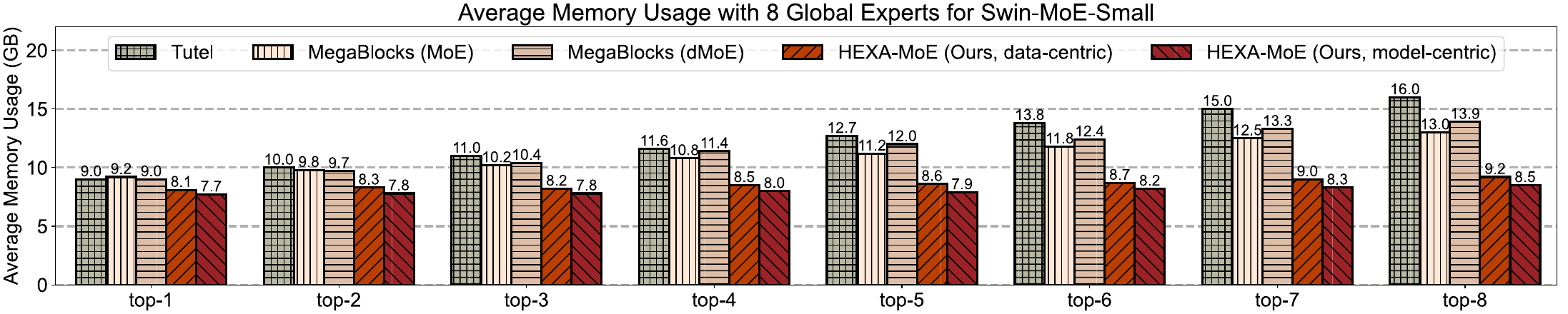}
    }
    \end{minipage}
    \vspace{-0.5cm}
    \caption{\textbf{Average memory usage for training Swin-Transformer-MoE models.} We take 8 global experts and examine all cases from top-1 to top-8 routings. Experiments are conducted on 2 homogeneous GPUs using automatic mixed precision in PyTorch. The batch size is set to 40 for all cases. We record the average GPU memory consumption (GB) on each device.
    }
    \label{fig:mem_charts}
    \vspace{-0.5cm}
\end{figure*}

\vspace{-0.3cm}
\subsection{Heterogeneous-aware Expert Allocation}\label{sec:hete}
\vspace{-0.1cm}
\paragraph{Workload Division for different configurations.} For data-centric, the workload mainly depends on local batch size, as it can be essentially viewed as data parallelism. For model-centric, the workload of each device mainly depends on the allocated sub-dimension of FFN intermediate size for each expert, as \textit{expert-specific} operators enable the implementation of tensor parallelism.
\vspace{-0.4cm}
\paragraph{Practical Expert Allocation on Heterogeneous Devices.} We propose a specialized expert allocation approach to utilize heterogeneous computing resources, by adjusting the workload of each device. Specifically, we have to first examine the computing capacity of each device by averaging its latency on a benchmark task with heavy computing, such as large matrix multiplication. Denoting the results as $\{t_i\}_{i=0}^{N-1}$ for $N$ GPUs. For data-centric, we assign different local batch size $\{B_i\}_{i=0}^{N-1}$ to different devices based on the examined latency, as illustrated in Equation~\ref{eq:bs-assign}:
\vspace{-0.3cm}
\begin{equation}
    \label{eq:bs-assign}
    B_i=\frac{1/t_i}{\sum_{j=0}^{N-1} 1/t_j}\cdot B_\textrm{global},
\end{equation}

\vspace{-0.5cm}
where we denote $B_\textrm{global}$ as the global batch size, \ie, $B_\textrm{global}=\sum_{j=0}^{N-1} B_i$. For model-centric, we denote the sub-dimension of hidden size for one MoE layer on each device as $\{h_i\}_{i=0}^{N-1}$ with total hidden size $H$. The assigned sub-dimension $h_i$ on device $i$ can be derived as Equation~\ref{eq:hid-assign}.
\vspace{-0.3cm}
\begin{equation}
    \label{eq:hid-assign}
    h_i=\frac{1/t_i}{\sum_{j=0}^{N-1} 1/t_j}\cdot H.
\end{equation}

\vspace{-0.5cm}
To ensure that both $\{B_i\}_{i=0}^{N-1}$ and $\{h_i\}_{i=0}^{N-1}$ are integers, we need to further round them up or down while making sure that the summation is exactly $B_\textrm{global}$ or $H$. 

\begin{table}[t]
    \centering
    \vspace{-0.2cm}
    \caption{\textbf{Details of the machines and GPUs used for both homogeneous and heterogeneous experiments.}}
    \scalebox{0.77}{
    \small
    \begin{tabular}{c|c|c|c}
        \toprule
         & Notation & Number \& Hardware Specs & Memory \\
        \cmidrule{1-4}
        \multirow{2}*{CPU} & $M_\textrm{homo}$ & $2\times$ Intel Xeon Platinum 8352V 2.10GHz & 1008 GB \\
        \cmidrule{2-4}
         & $M_\textrm{hete}$ & $2\times$ Intel Xeon Gold 6130 2.10GHz & 62.5 GB \\
        \cmidrule{1-4}
        \multirow{4}*{GPU} & $D$ (in $M_\textrm{homo}$) & $4\times$ NVIDIA GeForce RTX 4090 & 24 GB \\
        \cmidrule{2-4}
         & $D_0$ (in $M_\textrm{hete}$) & $1\times$ NVIDIA TITAN RTX & 24 GB \\
        \cmidrule{2-4}
         & $D_1$ (in $M_\textrm{hete}$) & $1\times$ NVIDIA GeForce RTX 2080 Ti & 11 GB \\
        \bottomrule
    \end{tabular}}
    \vspace{-0.4cm}
    \label{tab:clusters}
\end{table}

\vspace{-0.2cm}
\section{Experiments}
\vspace{-0.2cm}
\subsection{Experimental Setup}
\vspace{-0.2cm}
We implement all the experiments using 2 machines, namely $M_\textrm{homo}$ and $M_\textrm{hete}$. Specifically, $M_\textrm{homo}$ is composed of 4 homogeneous GPUs, while $M_\textrm{hete}$ contains 2 heterogeneous GPUs. Details for the machines and GPUs are provided in Table~\ref{tab:clusters}. We take the training process of Swin-Transformer-MoE as the benchmark for all the experiments, including memory analysis, latency analysis, heterogeneous computing analysis, and ablation studies. Apart from heterogeneous computing analysis, all the experiments are conducted on $M_\textrm{homo}$. We also adopt both the Small and Base scales for the Swin-MoE model, following Tutel~\cite{hwang2023tutel}. Experiments are conducted with PyTorch, taking \texttt{nccl} as the communication backend. Batch size is recorded as the workload of single device, and automatic mixed precision is used for training. \texttt{atomicAdd} is employed for \texttt{HEXA-MoE} to aggregate the expert computing results for each token.

\begin{figure*}[ht]
\centering
\begin{minipage}{1.0\linewidth}
    \centering
    \subfigure{
        \includegraphics[width=0.98\linewidth,height=1.4in]{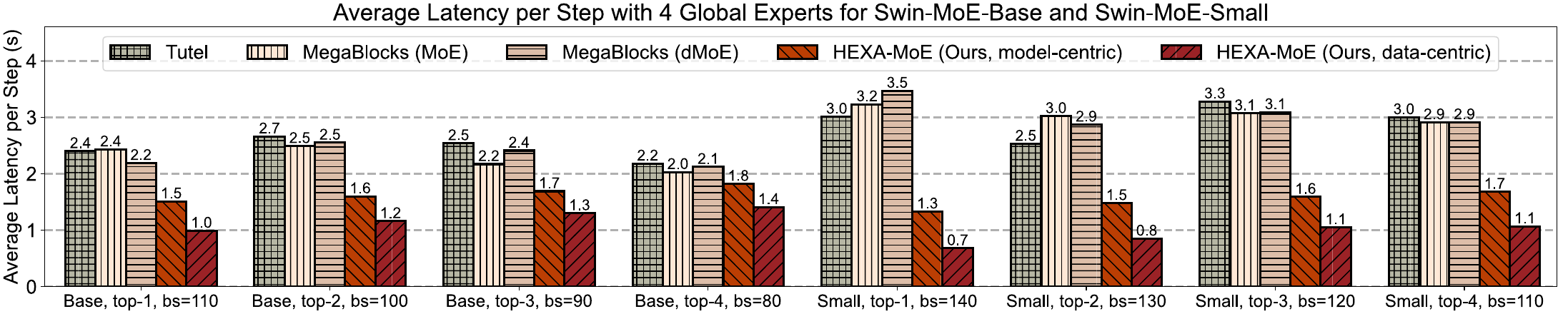}	
    }
\end{minipage}
\vspace{-0.7cm}
\caption{\textbf{Average latency for training Swin-Transformer-MoE models.} Experiments are conducted on 4 homogeneous GPUs with 4 global experts. We set different batch sizes for different models under different routing strategies to maximize the utilization of GPU memory. We record the average latency for one step (s) during training with 2k steps in total.}
\label{Fig:latency-charts}
\vspace{-0.7cm}
\end{figure*}

\vspace{-0.2cm}
\subsection{Experimental Results}
\begin{figure}[t]
\centering
\vspace{0.2cm}
\begin{minipage}{1.0\linewidth}
    \centering
    \subfigure{
        \includegraphics[width=0.98\linewidth,height=1.2in]{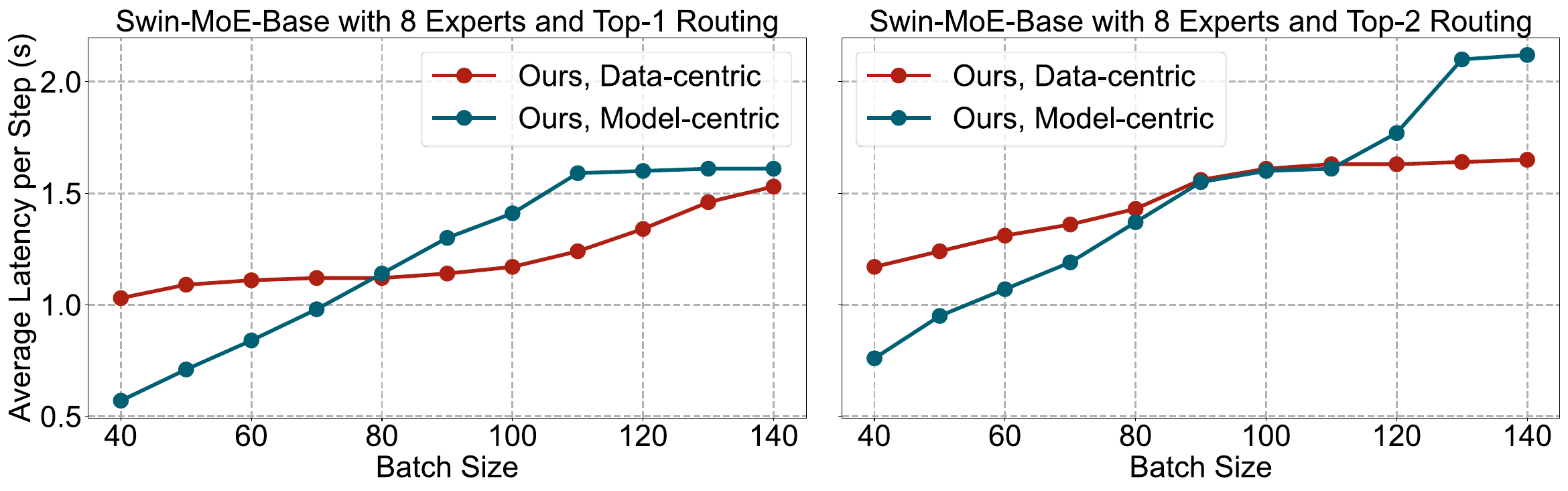}	
    }
\end{minipage}
\vspace{-0.7cm}
\caption{\textbf{Latency comparison with data-centric and model-centric configurations.} We record the average latency per step with 2k steps in total. The data-centric setting presents a more gentle trend when scaling workload.}
\label{Fig:data-vs-model}
\vspace{-0.7cm}
\end{figure}

\vspace{-0.2cm}
\paragraph{Memory Analysis on Homogeneous Devices.} We analyze the memory footprint for different MoE libraries on homogeneous devices with 8 global experts, and examine from top-$1$ to top-$8$ routing. Results are visualized in Figure~\ref{fig:mem_charts}. \texttt{HEXA-MoE} can reduce $10\%$-$48\%$ memory footprint compared to Tutel~\cite{hwang2023tutel} and MegaBlocks~\cite{gale2023megablocks}, and the memory footprint for model-centric is slightly less than data-centric owing to the pipeline-shared cache. From top-$1$ to top-$k$ routing, the memory footprint increase of \texttt{HEXA-MoE} is more gentle than others, owing to the introduction of memory optimization. Memory saving mainly benefits from the in-place design of \textit{expert-specific} operators and memory optimization. 
\vspace{-0.4cm}
\paragraph{Latency Analysis on Homogeneous Devices.} We analyze the average latency for one training step, and present the results in Figure~\ref{Fig:latency-charts}. To fully utilize the GPU HBM, we set different batch sizes for different routing configurations and different model scales. \texttt{HEXA-MoE} achieves $0.5$-$4.3$$\times$ speed up compared to Tutel~\cite{hwang2023tutel} and MegaBlocks~\cite{gale2023megablocks}. Since the scale of workload is larger than MoE parameters in our experiments, the average latency under a data-centric setting is less than model-centric in these cases, and the advantage of \texttt{HEXA-MoE} turns more obvious while enlarging batch size. To better visualize the latency comparison between data-centric and model-centric configurations, we present the latency for the Swin-MoE-Base model under different batch sizes in Figure~\ref{Fig:data-vs-model}. When the workload is relatively small, a model-centric setting is more efficient, while a data-centric setting becomes more efficient with a relatively large workload. Experiments are all conducted on the homogeneous machine with automatic mixed precision in PyTorch. 

\begin{figure*}[ht]
\centering
\begin{minipage}{1.0\linewidth}
    \centering
    \subfigure{
        \label{subfig:data-centric-hete}
        \includegraphics[width=0.99\linewidth,height=1.3in]{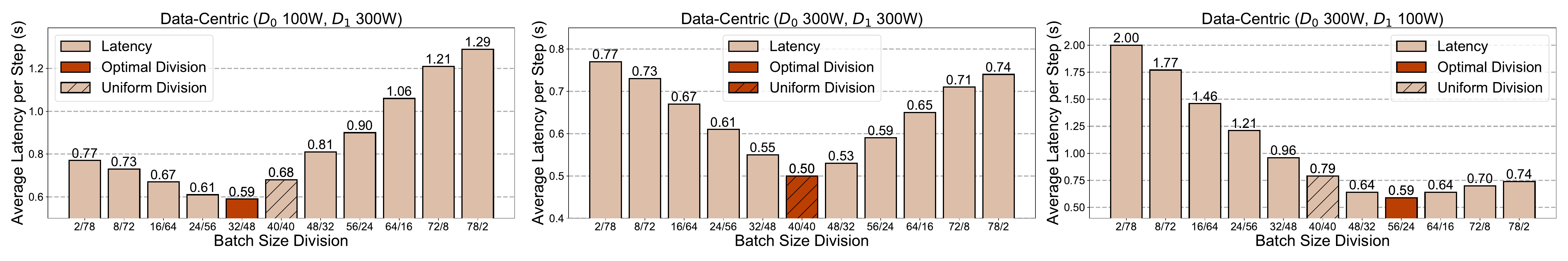}	
    }
\end{minipage}\vspace{-0.4cm}
\begin{minipage}{1.0\linewidth}
    \centering
    \subfigure{
        \label{subfig:model-centric-hete}
        \includegraphics[width=0.99\linewidth,height=1.3in]{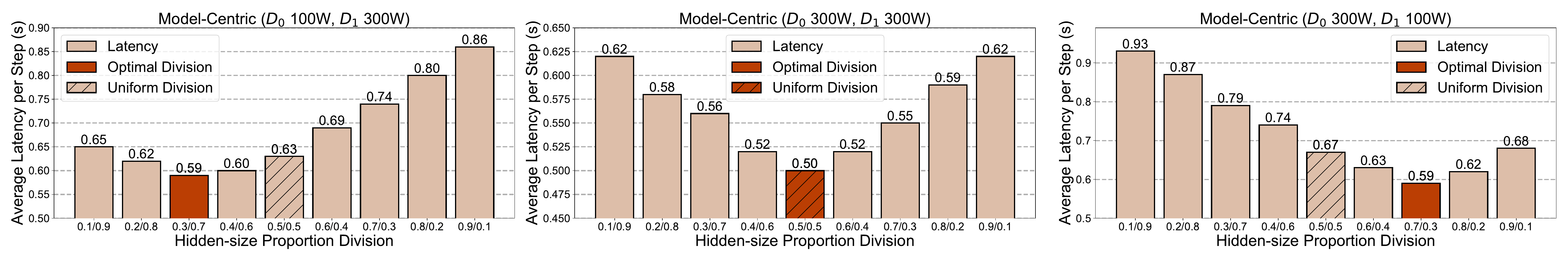}
    }
\end{minipage}
\vspace{-0.8cm}
\caption{\textbf{Latency analysis on heterogeneous devices with both data-centric and model-centric configuration.} Experiments are all conducted with Swin-MoE-Small model using 4 global experts and top-2 routing. The average latency falls minimal when the division proportion is close to the computing capacity proportion in each case.}
\label{Fig:hete-charts}
\vspace{-0.5cm}
\end{figure*}

\vspace{-0.4cm}
\paragraph{Heterogeneous Experiments.} We demonstrate the heterogeneous awareness of our \texttt{HEXA-MoE} via experiments on different devices under both data-centric and model-centric configurations, and visualize the average latency with different division strategies in Figure~\ref{Fig:hete-charts}. Since the $2$ GPUs in our heterogeneous machine have similar computing capacities, we also adjust the power limit for each device. We first examine the computing capacity for these experimental settings, and provide the results in Table~\ref{tab:capacity}. The details of the proxy task are provided in Appendix~\ref{appendix:details}. 

In Figure~\ref{subfig:data-centric-hete}, we adjust the batch size on each device under data-centric configuration. Essentially the optimal proportion is very close to the examined computing capacity proportion, and it has a certain deviation from uniform division when the power constraints of the two devices are not equal, while the optimal division is also uniform division when the power constraints are equal. Specifically, when setting the power constraint for $D_0$ to be 100W and $D_1$ to be 300W, employing our heterogeneous-aware allocation can make the average latency to be 13.2\% lower than naive division, and when setting $D_0$ to be 300W and $D_1$ to be 100W, our approach can reduce the average latency by 25.3\%. 

In Figure~\ref{subfig:model-centric-hete}, we adjust the allocated sub-dimension proportion of the hidden size for each MoE layer on each device. Adapting the sub-dimension proportion to be close to the computing capacity proportion of heterogeneous devices can also substantially minimize average latency, similar to the data-centric setting. Specifically, when setting the power constraint of $D_0$ to be 100W and $D_1$ to be 300W, our approach can achieve a 6.3\% reduction on average latency compared to uniform division, and when setting $D_0$ to be 300W and $D_1$ to be 100W, we can reduce it by 11.9\%. Although the reduction is not as significant as a data-centric setting, we can still achieve a relative speed-up. These demonstrate that our \texttt{HEXA-MoE} can essentially maximize the utilization of heterogeneous computing resources.

\begin{table}[t]
    \centering
    \caption{\textbf{Examining computing capacity proportion for heterogeneous devices.} $P$, $T$ and $R$ denote power constraint (W), average latency (s), and computing capacity proportion.}
    \scalebox{0.82}{
    \small
    \begin{tabular}{c|c|c|c|c|c|c|c|c|c}
        \toprule
         & \multicolumn{3}{c|}{Case 1} & \multicolumn{3}{c|}{Case 2} & \multicolumn{3}{c}{Case 3} \\
        \cmidrule{2-10}
         & $P$ & $T$ & $R$ & $P$ & $T$ & $R$ & $P$ & $T$ & $R$ \\
        \cmidrule{1-10}
        $D_0$ & 100 & 4.58 & 0.40 & 300 & 3.20 & 0.50 & 300 & 3.28 & 0.74 \\
        \cmidrule{1-10}
        $D_1$ & 300 & 3.06 & 0.60 & 300 & 3.18 & 0.50 & 100 & 9.42 & 0.26 \\
        \bottomrule
    \end{tabular}}
    \label{tab:capacity}
    \vspace{-0.4cm}
\end{table}

\vspace{-0.3cm}
\subsection{Ablation Studies}
\begin{figure}[t]
\centering
\begin{minipage}{0.98\linewidth}
    \centering
    \subfigure{
        \label{fig:mem_breakdown}
        \includegraphics[width=0.96\linewidth,height=1.8in]{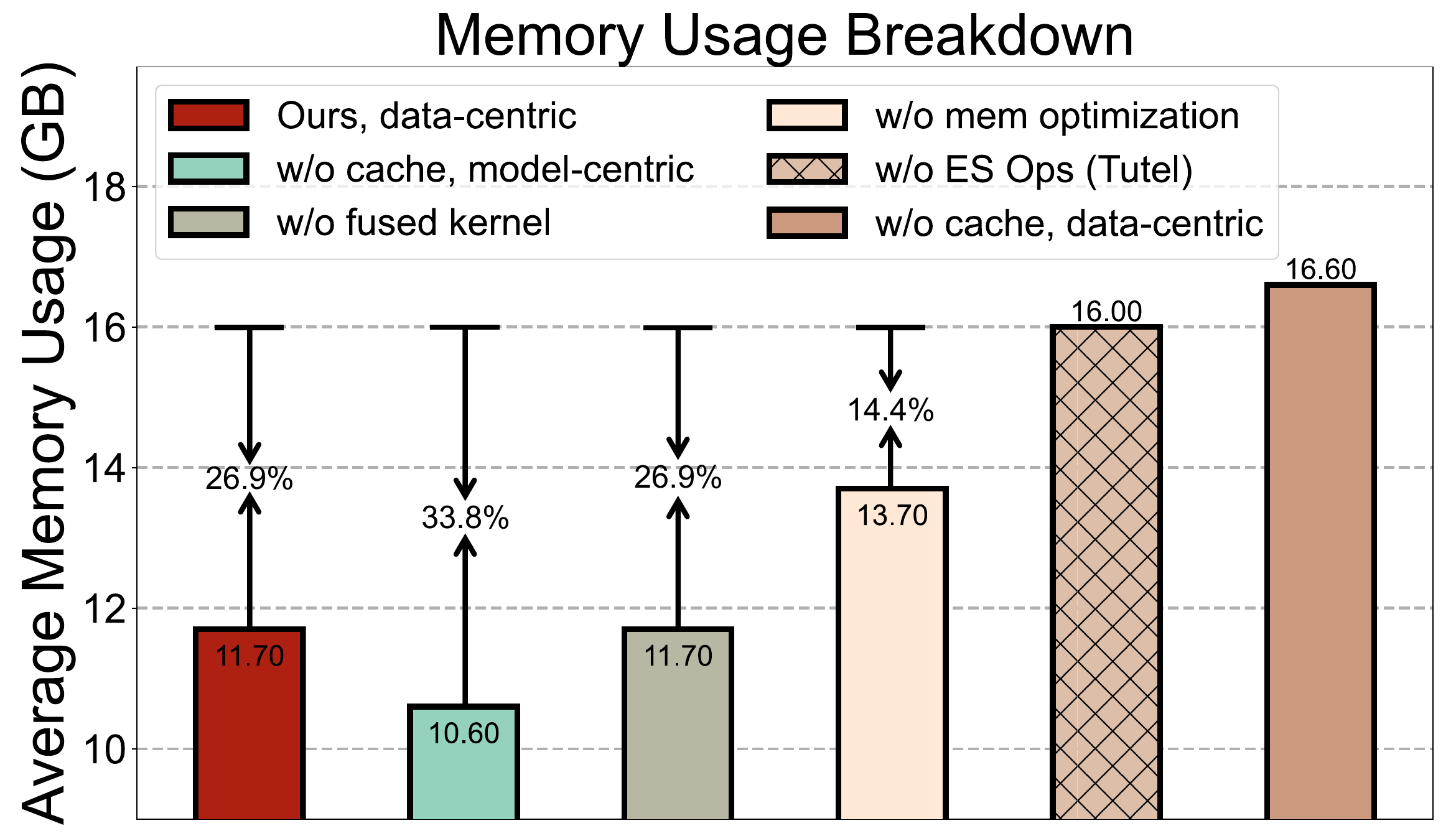}
    }
\end{minipage}\\
\vspace{-0.26cm}
\begin{minipage}{0.98\linewidth}
    \centering
    \subfigure{
        \label{fig:latency_breakdown}
        \includegraphics[width=0.96\linewidth,height=1.8in]{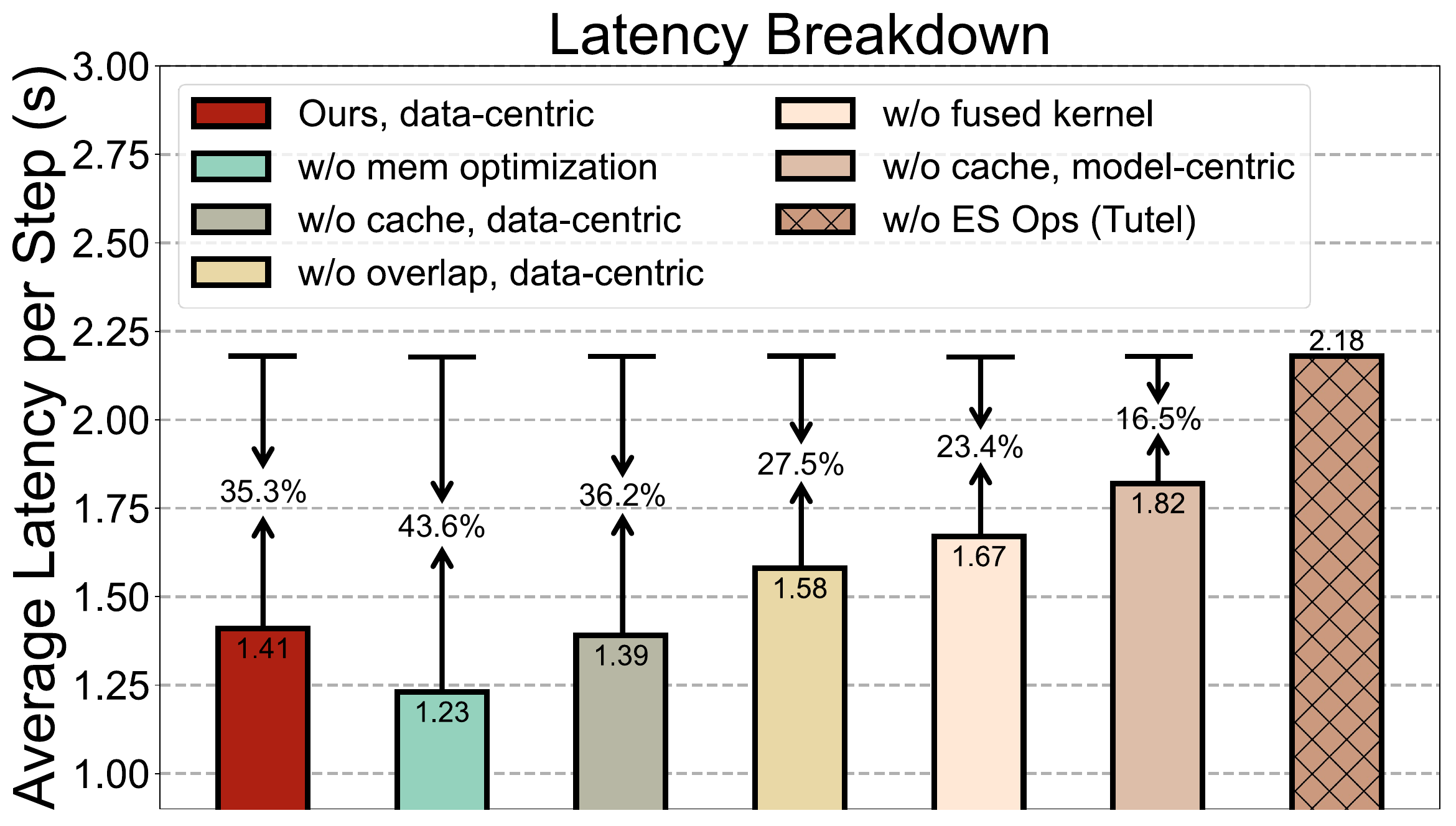}
    }
\end{minipage}
\vspace{-0.6cm}
\caption{\textbf{Memory and latency breakdown}. We analyze the effectiveness of our proposed pipeline-shared cache, fused kernel, and memory optimization. We take Tutel as a baseline and visualize the differences with it.}
\label{Fig:breakdown}
\vspace{-0.7cm}
\end{figure}

\vspace{-0.2cm}
We examine the effectiveness of each component for \texttt{HEXA-MoE} via measuring the impact of expert-specific operators, pipeline-shared cache, fused kernel, data- \& model-centric and memory optimization, respectively. $2$ metrics are evaluated, including average latency and memory footprint. We take distributed training of Swin-MoE-Base model on homogeneous devices as benchmark for ablation studies. Since dispatch \& combine operations are inevitable if we dispense with our expert-specific operators, we directly take the performance of Tutel in this case.
\vspace{-0.4cm}
\paragraph{Memory Footprint Breakdown.} We take 8 experts, top-4 routing, and batch size 40 as an example to break down memory footprint, and visualize the results in Figure~\ref{fig:mem_breakdown}. We can draw some insights from it:
\vspace{-0.4cm}
\begin{enumerate}
    \item [$\star$] The employment of pipeline-shared cache slightly increases memory footprint, while expert-specific fused kernel has no impact.
    \vspace{-0.3cm}
    \item [$\star$] Pipeline-shared cache can effectively reduce memory footprint under a data-centric setting, since the memory footprint would surpass Tutel without it.
    \vspace{-0.3cm}
    \item [$\star$] \texttt{HEXA-MoE} can reduce memory footprint compared to baseline even without memory optimization, and optimizing it can further reduce memory footprint.
\end{enumerate}

\vspace{-0.6cm}
\paragraph{Latency Breakdown.} We set 4 experts, top-4 routing, and batch size 80 to break down training latency, and visualize results in Figure~\ref{fig:latency_breakdown}. We can draw some insights as:
\vspace{-0.4cm}
\begin{enumerate}
    \item [$\star$] The employment of expert-specific fused kernel, data-centric setting, and communication-computation overlap can effectively reduce latency. Meanwhile, the combination of data-centric setting and communication-computation overlap can further empower our \texttt{HEXA-MoE} with notably reduced latency.
    \vspace{-0.6cm}
    \item [$\star$] Expert-specific operators play a significant part in speeding up MoE computing since the latency for Tutel is much longer than our model-centric setting.
    \vspace{-0.2cm}
    \item [$\star$] Although pipeline-shared cache and memory optimization can slightly increase latency, they can also lead to significant reduction on memory footprint, as demonstrated in memory footprint breakdown.
\end{enumerate}

\vspace{-0.4cm}
\section{Conclusion}
We introduce \texttt{HEXA-MoE}, a completely static MoE training framework with minimal memory footprint and heterogeneous-awareness, materialized by the proposed \textit{expert-specific} operators and tensor parallelism. We provide specialized GPU kernels, which can both \underline{save memory footprint} and \underline{reduce overall latency}. For different workload scales, we provide data- and model-centric configurations for enhanced efficiency, and propose a memory-efficient communication-computation overlapping scheme to overcome the shortcomings of previous work. Homogeneous experiments show that \texttt{HEXA-MoE} can save $10\%$-$48\%$ memory footprint while achieving $0.5$-$4.3$$\times$ speed up compared to state-of-the-art MoE libraries. Heterogeneous experiments show that \texttt{HEXA-MoE} can substantially minimize runtime and better utilize global computing resources by employing optimal parallel configuration. \texttt{HEXA-MoE} endows MoE computing with an in-place manner, where routing choice would yield almost no impact on hardware workload, which may inspire further research on algorithm designs of MoE.

\section*{Impact Statement}

This paper aims to improve the computation efficiency for training MoE models. The efficiency advantage of such models might help democratize access of large-scale foundation models. On the other hand, whether such new algorithm would affect known issues such as biased and harmful outputs of language models remains an unexplored research question.



\nocite{*} 


\newpage
\appendix
\onecolumn
\section{Principles of CUDA Programming}

\begin{figure}[H]
\centering
\begin{minipage}{0.49\linewidth}
\centering
\subfigure[Thread hierarchy for CUDA programming.]{
    \label{fig:thread-hierarchy}
    \includegraphics[width=0.98\linewidth,height=1.3in]{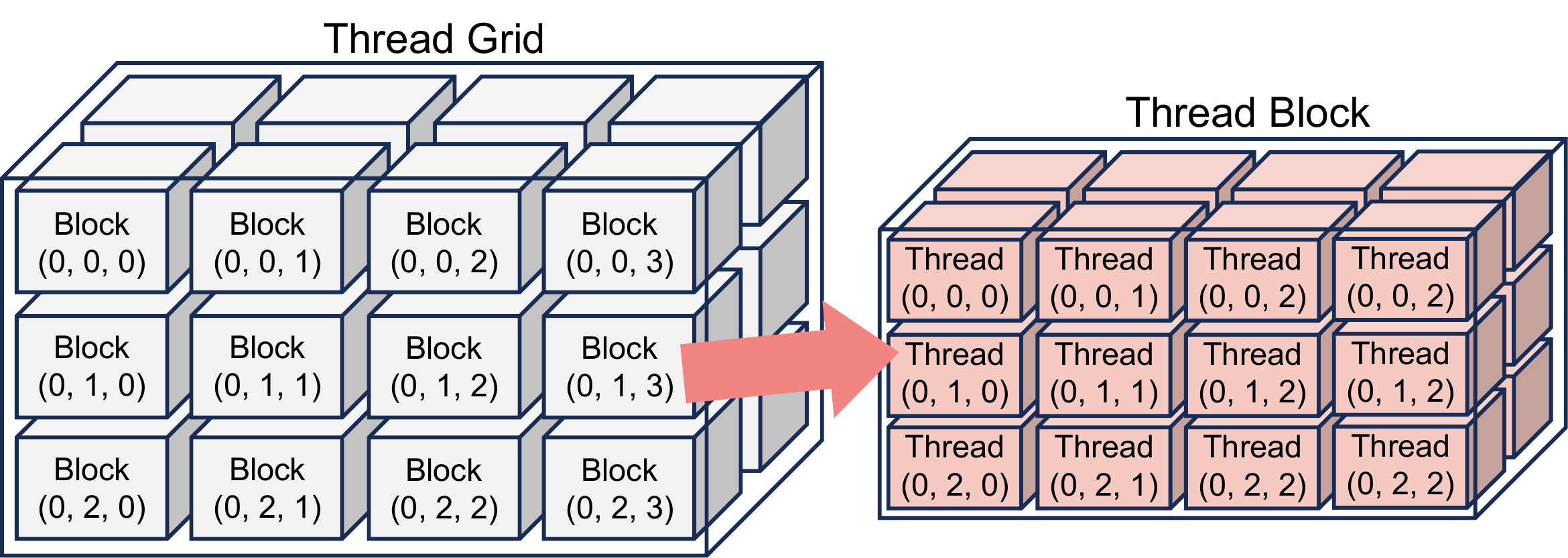}	
}
\end{minipage}
\begin{minipage}{0.49\linewidth}
\subfigure[Memory hierarchy for a GPU.]{
    \label{fig:memory-hierarchy}
    \includegraphics[width=0.98\linewidth,height=1.3in]{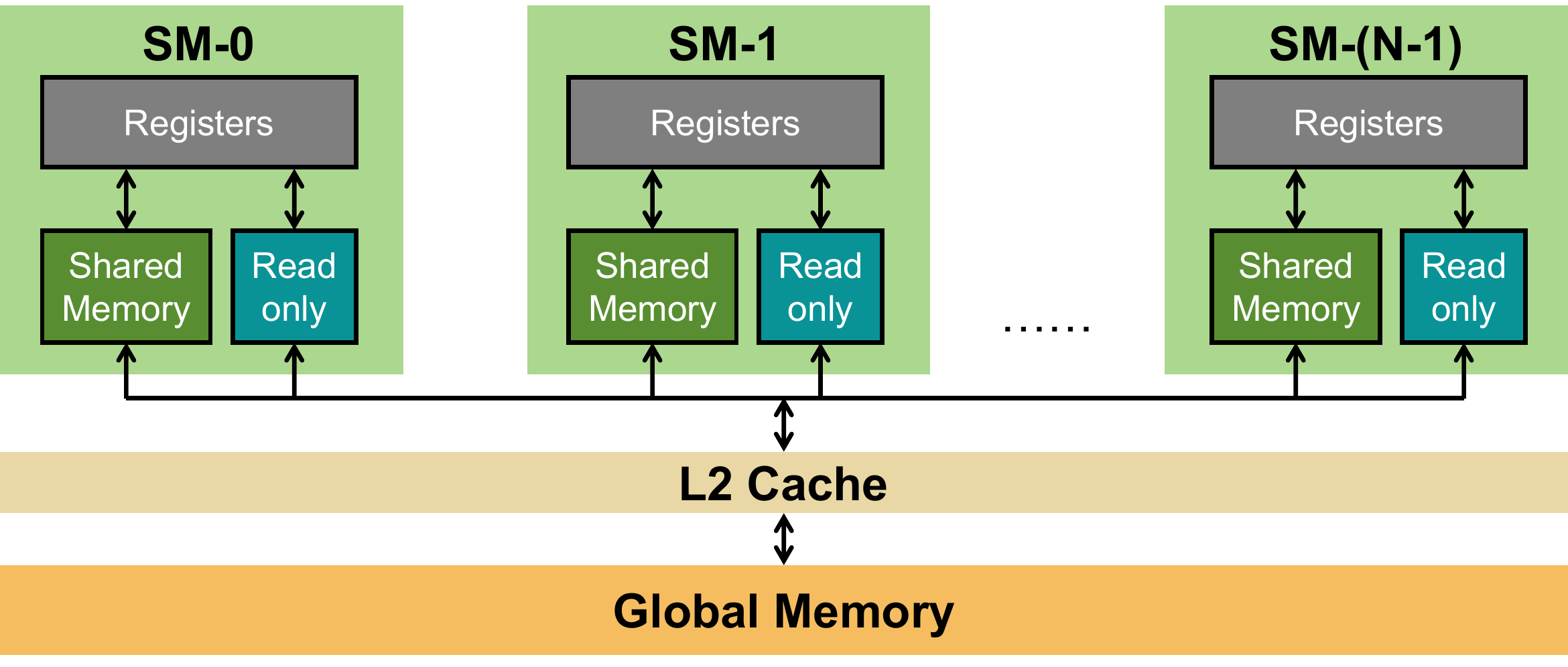}
}
\end{minipage}
\vspace{-0.4cm}
\caption{\textbf{Thread hierarchy for CUDA programming and memory hierarchy for a typical NVIDIA GPU.}}
\label{fig:hierarchy}
\vspace{-0.2cm}
\end{figure}

\paragraph{Introduction to CUDA Programming:} A GPU program can be viewed as a thread grid in Figure~\ref{fig:thread-hierarchy}, which is $3$-dimensional and contains massive thread blocks. A thread block is also $3$-dimensional and contains up to $1024$ threads. Computation is mapped to these threads and implemented in parallel. Physically a GPU is composed of massive streaming multiprocessors (SM), where we show the memory hierarchy in Figure~\ref{fig:memory-hierarchy}. Each SM loads data from global memory to its registers for parallel computing, and caching with shared memory can be faster. An SM contains many processing units including CUDA core and Tensor core. The former is used for general parallel computing, while the latter is specialized for matrix multiplication with mixed precision, which is first introduced in NVIDIA Volta architecture~\cite{choquette2018volta}. During execution, an SM is given one or more thread blocks, which are partitioned into warps and each warp gets scheduled by a warp scheduler. 

\vspace{-0.4cm}
\paragraph{Mathematical Symbols:} We provide the description of mathematical symbols used in this paper in Table~\ref{tab:symbols}, including variables in MoE computing and constant values in CUDA programming. To utilize Tensor core for faster matrix processing, we have to call the \texttt{nvcuda::wmma} (warped matrix multiplication and add) interface to compute a fixed-sized matrix multiplication such as $16$$\times$$16$$\times16$ in a single warp, therefore the number of threads in a block has to be divisible by the constant value \texttt{WARP}. In the CUDA implementation of our method, a thread block loads \texttt{BLK} tokens each time from global memory to shared memory to conduct the parallelized computation.

\begin{table}[H]
    \centering
    \caption{\textbf{Description of symbols.}}
    \scalebox{0.9}{
    \small
    \begin{tabular}{c|c}
        \toprule
        Symbol & Description \\
        \cmidrule{1-2}
        $E$ & Number of global experts. \\
        \cmidrule{1-2}
        $N$ & Number of the input tokens. \\
        \cmidrule{1-2}
        $\mathcal{F}$ & Activation function between the two MLPs. \\
        \cmidrule{1-2}
        $\mathcal{F}^{\prime}$ & Element-wise differential for $\mathcal{F}$. \\
        \cmidrule{1-2}
        $\odot$ & Element-wise product. \\
        \cmidrule{1-2}
        $D_i, D_o$ & Input and output size of the FFN. \\
        \cmidrule{1-2}
        $\boldsymbol{x}$ & Data batch with $N$ tokens. \\
        \cmidrule{1-2}
        $\boldsymbol{x}^e, N_e$ & Tokens in $\boldsymbol{x}$ routed to expert $e$ with number $N_e$. \\
        \cmidrule{1-2}
        $\{\mathcal{R}_i(\boldsymbol{x})\}_{i=0}^{k-1}$ & Routing choice for $\boldsymbol{x}$ under top-$k$ routing. \\
        \cmidrule{1-2}
        $\boldsymbol{W_1}, \boldsymbol{W_2}$ & Weights for 2 MLPs, shaped as $(E, D_1, D_2)$. \\
        \cmidrule{1-2}
        $\boldsymbol{b_1}, \boldsymbol{b_2}$ & Biases of the two MLPs, shaped as $(E, D)$. \\
        \cmidrule{1-2}
        $\texttt{BLK}$ & Block size for expert-specific operators. \\
        \cmidrule{1-2}
        $\texttt{WARP}$ & Warp size in a thread block. \\
        \cmidrule{1-2}
        $\texttt{TIMES}$ & A thread block has         $\texttt{WARP}$$\times$$\texttt{TIMES}$ threads. \\
        \bottomrule
    \end{tabular}}
    \label{tab:symbols}
    \vspace{-0.2cm}
\end{table}

\section{Algorithm Details}\label{appendix:algorithm}

\begin{algorithm}[H]
    \caption{Constructing re-index vector}
    \label{alg:re-index}
    \begin{algorithmic}
        \State {\bfseries Input:} Routing choice $\mathcal{R}$ with shape $(N,)$.
        \State {\bfseries Initialize:} Tensor $ctr$ with shape $(E,)$ initialized with 0.
        \State {\bfseries {Parallel for} $i=0$ {\bfseries to} $N-1$ {\bfseries do}}
        \State {\quad} \texttt{atomicAdd}($ctr[\mathcal{R}[i]]$, 1)
        \State {\bfseries end for}

        \State {\bfseries {Parallel for} $i=0$ {\bfseries to} $E-1$ {\bfseries do}}
        \State {\quad} $ctr[i]=\texttt{BLK}\cdot\lfloor ctr[i]\ /\ \texttt{BLK}\rfloor$
        \State {\bfseries end for}

        $N^{\prime}=\sum\limits_{i=0}^{E-1}ctr[i]$
        
        \State {\bfseries Initialize:} Tensor $v$ with shape $(N^{\prime},)$ initialized with -1, and tensor $idx$ with shape $(1+E,)$ initialized with 0.

        \State {\bfseries {Parallel for} $i=0$ {\bfseries to} $E-1$ {\bfseries do}}
        \State {\quad} $ctr[i]=\texttt{BLK}\cdot\lfloor ctr[i]\ /\ \texttt{BLK}\rfloor$
        \State {\bfseries end for}

        \State {\bfseries {for} $i=1$ {\bfseries to} $E$ {\bfseries do}}
        \State {\quad} $idx[i]=idx[i-1]+ctr[i-1]$
        \State {\bfseries end for}

        \State {\bfseries Copy:} $idx\_=idx$

        \State {\bfseries {Parallel for} $i=0$ {\bfseries to} $N-1$ {\bfseries do}}
        \State {\quad} $pos=\texttt{atomicAdd}(idx[\mathcal{R}[i]], 1)$
        \State {\quad} $v[pos]=i$
        \State {\bfseries end for}
        
        \State {\bfseries Output:} Tensor $v$ and $idx\_$
    \end{algorithmic}
\end{algorithm}

\begin{algorithm}[H]
    \caption{Expert-specific matrix multiplication}
    \label{alg:esmm}
    \begin{algorithmic}
    \State {\bfseries Input:} Routing choice $\mathcal{R}$ with shape $(N,)$, vector $v$ with length $N^{\prime}$, input tokens $x$ with shape $(N, D_1)$, weights $w$ with shape $(E, D_1, D_2)$ and bias $b$ with shape $(E, D_2)$
    \State {\bfseries Initialize:} Output tokens $y$ with shape $(N, D_2)$ initialized with 0.

    \State {\bfseries {Parallel for} $i$ {\bfseries in range}} $(0, N^{\prime}, \texttt{BLK}) \ \textbf{do}$
    \State {\quad \bfseries {Parallel for} $j$ {\bfseries in range}} $(0, D_2, \texttt{BLK}) \ \textbf{do}$
    \State {\quad\quad} $exp=\mathcal{R}[v[i]]$
    \State {\quad\quad \bfseries {Initialize}} zero tensor $c$ with shape $(\texttt{BLK}, \texttt{BLK})$
    \State {\quad\quad \bfseries {load}} $b_{sub}=b[exp, j:j+\texttt{BLK}]$
    \State {\quad\quad} $c=b_{sub}.\text{repeat}(\texttt{BLK, 1})$
    \State {\quad\quad \bfseries {for} $k$ {\bfseries in range}} $(0, D_1, \texttt{BLK}) \ \textbf{do}$
    \State {\quad\quad\quad \bfseries {Initialize}} zero tensor $x_{sub}$ with shape $(\texttt{BLK}, \texttt{BLK})$
    \State {\quad\quad\quad \bfseries {Parallel for} $t=0$ {\bfseries to}} $\texttt{BLK} \ \textbf{do}$
    \State {\quad\quad\quad\quad \bfseries {load}} $x_{sub}[t]=x[v[i+t], k:k+\texttt{BLK}]$
    \State {\quad\quad\quad \bfseries {end for}}
    \State {\quad\quad\quad \bfseries {load}} $w_{sub}=w[exp, k:k+\texttt{BLK}, j:j+\texttt{BLK}]$
    \State {\quad\quad\quad} $c=c+x_{sub}\cdot w_{sub}$
    \State {\quad\quad \bfseries end for}
    \State {\quad\quad \bfseries {Parallel for} $t=0$ {\bfseries to}} $\texttt{BLK} \ \textbf{do}$
    \State {\quad\quad\quad \bfseries {Write back:}} $y[v[i+t], j:j+\texttt{BLK}]=c[t]$
    \State {\quad\quad \bfseries {end for}}
    \State {\quad \bfseries end for}
    \State {\bfseries end for}
    
    \State {\bfseries Output:} Tensor $y$
    \end{algorithmic}
\end{algorithm}

\begin{algorithm}[H]
    \caption{Expert-specific summation}
    \label{alg:ess}
    \begin{algorithmic}
    \State {\bfseries Input:} Routing choice $\mathcal{R}$ with shape $(N,)$, vector $v$ with length $N^{\prime}$, vector $idx$ with length $1+E$ and input tokens $x$ with shape $(N, D)$
    \State {\bfseries Initialize:} Output tokens $y$ with shape $(E, D)$ initialized with 0.

    \State {\bfseries {Parallel for} $i=0$ {\bfseries to} $E-1$ {\bfseries do}}
    \State {\quad \bfseries {Parallel for} $j$ {\bfseries in range}} $(0, D, \texttt{BLK}) \ \textbf{do}$
    \State {\quad\quad} $exp=\mathcal{R}[v[idx[i]]]$
    \State {\quad\quad \bfseries {Initialize}} zero tensor $c$ with shape $(1, \texttt{BLK})$
    \State {\quad\quad \bfseries {Initialize}} zero tensor $x_{sub}$ with shape $(\texttt{BLK}, \texttt{BLK})$
    \State {\quad\quad \bfseries {for} $k$ {\bfseries in range}} $(idx[i], idx[i+1], \texttt{BLK}) \ \textbf{do}$
    \State {\quad\quad\quad \bfseries {Parallel for} $t=0$ {\bfseries to}} $\texttt{BLK} \ \textbf{do}$
    \State {\quad\quad\quad\quad \bfseries {load}} $x_{sub}[t]=x[v[k+t], j:j+\texttt{BLK}]$
    \State {\quad\quad\quad \bfseries {end for}}
    \State {\quad\quad\quad} $c=c+\sum\limits_{t=0}^{\texttt{BLK}-1}x_{sub}[t]$
    \State {\quad\quad \bfseries {end for}}
    \State {\quad \bfseries Write back:} $y[exp, j:j+\texttt{BLK}]=c$
    \State {\quad \bfseries end for}
    \State {\bfseries end for}

    \State {\bfseries Output:} Tensor $y$
    \end{algorithmic}
\end{algorithm}

\begin{algorithm}[H]
    \caption{Expert-specific transposed matrix multiplication}
    \label{alg:estmm}
    \begin{algorithmic}
    \State {\bfseries Input:} Routing choice $\mathcal{R}$ with shape $(N,)$, vector $v$ with length $N^{\prime}$, vector $idx$ with length $1+E$, the first token batch $x_1$ with shape $(N, D_1)$ and the second token batch $x_2$ with shape $(N, D_2)$
    \State {\bfseries Initialize:} Output $y$ with shape $(E, D_1, D_2)$ initialized with 0.

    \State {\bfseries {Parallel for} $i=0$ {\bfseries to} $E-1$ {\bfseries do}}
    \State {\quad \bfseries {Parallel for} $m$ {\bfseries in range}} $(0, D_1, \texttt{BLK}) \ \textbf{do}$
    \State {\quad\quad \bfseries {Parallel for} $n$ {\bfseries in range}} $(0, D_2, \texttt{BLK}) \ \textbf{do}$
    \State {\quad\quad\quad} $exp=\mathcal{R}[v[idx[i]]]$
    \State {\quad\quad\quad \bfseries {Initialize}} zero tensor $c$ with shape $(\texttt{BLK}, \texttt{BLK})$
    \State {\quad\quad\quad \bfseries {Initialize}} zero tensor $x_{sub}^1$ with shape $(\texttt{BLK}, \texttt{BLK})$
    \State {\quad\quad\quad \bfseries {Initialize}} zero tensor $x_{sub}^2$ with shape $(\texttt{BLK}, \texttt{BLK})$
    \State {\quad\quad\quad \bfseries {for} $k$ {\bfseries in range}} $(idx[i], idx[i+1], \texttt{BLK}) \ \textbf{do}$
    \State {\quad\quad\quad\quad \bfseries {Parallel for} $t=0$ {\bfseries to}} $\texttt{BLK} \ \textbf{do}$
    \State {\quad\quad\quad\quad\quad \bfseries {load}} $x_{sub}^1[t]=x_1[v[k+t], m:m+\texttt{BLK}]$
    \State {\quad\quad\quad\quad\quad \bfseries {load}} $x_{sub}^2[t]=x_2[v[k+t], n:n+\texttt{BLK}]$
    \State {\quad\quad\quad\quad \bfseries {end for}}
    \State {\quad\quad\quad\quad} $c=c+x_{sub}^1.\text{transpose()}\cdot x_{sub}^2$
    \State {\quad\quad\quad \bfseries end for}
    \State {\quad\quad\quad \bfseries Write back in parallel:} 
    \State {\quad\quad\quad\quad} $y[exp, m:m+\texttt{BLK}, n:n+\texttt{BLK}]=c$
    \State {\quad\quad \bfseries end for}
    \State {\quad \bfseries end for}
    \State {\bfseries end for}
    
    \State {\bfseries Output:} Tensor $y$
    \end{algorithmic}
\end{algorithm}

We present the algorithm details for re-index vector construction as well as the expert-specific operators, and taking top-1 routing as an example for illustration, shown in Algorithm~\ref{alg:re-index}, \ref{alg:esmm}, \ref{alg:ess} and \ref{alg:estmm}. In Algorithm~\ref{alg:re-index}, we re-arrange the routing choice vector $\mathcal{R}$ into re-indexed token vector $v$, along with the token index starting vector $idx$, which satisfies $idx[0]=0$ and $idx[E]=N^{\prime}$. We provide the length and range for the vectors in Table~\ref{tab:auxiliary}.

\begin{table}[ht]
    \centering
    \caption{\textbf{Explanation for the auxiliary vectors.}}
    \small
    \begin{tabular}{c|c|c}
        \toprule
        vector & length & range \\
        \midrule
        $\mathcal{R}$ & $N$ & $0\leq\mathcal{R}[i]<E$ for $0\leq i<N$ \\
        \midrule
        $v$ & $N^{\prime}$ & $0\leq v[i]<N$ for $0\leq i<N^{\prime}$ \\
        \midrule
        $idx$ & $1+E$ & $0\leq idx[i]<N^{\prime}$ for $0\leq i<1+E$ \\
        \bottomrule
    \end{tabular}
    \label{tab:auxiliary}
    \vspace{-0.2cm}
\end{table}

In Algorithm~\ref{alg:esmm}, \ref{alg:ess} and \ref{alg:estmm}, we assume that the feature dimension for each tensor is all divisible by $\texttt{BLK}$. Notice that when accessing the re-index vector $v$, we may get value -1, since the workload for each expert is dynamic. In this case, we would skip this index, and remain zero for the temporary loading variable.

We also provide the formulation for top-$k$ routing with our expert-specific operators, and take a comparison with top-$1$ routing. We denote the routing choice for top-$k$ as $\{\mathcal{R}_i(\boldsymbol{x})\}_{i=0}^{k-1}$, and other symbols keep consistent with Figure~\ref{fig:formula}. We present the comparison on formulations for MoE forward and backward propagation in Table~\ref{tab:top-k}.

\begin{table*}[ht]
    \centering
    \caption{\textbf{Comparison between top-$1$ and top-$k$ routing for the formulation of MoE forward and backward with our expert-specific operators.}}
    \scalebox{0.96}{
        \small
        \begin{tabular}{c|c|c|c|c}
            \toprule
            Stage & Notation & Layer & \makecell{Expert-Specific Formulation \\ (top-$1$ routing)} & \makecell{Expert-Specific Formulation \\ (top-$k$ routing)} \\
            \cmidrule{1-5}
            \multirow{4}{*}{Forward} & $\circled{1}$ & 1st MLP  & $\boldsymbol{y}_1=\textit{ESMM}(\boldsymbol{x}, \boldsymbol{W}_1, \boldsymbol{b}_1, \mathcal{R}(\boldsymbol{x}))$ & $\{\boldsymbol{y}_1^i\}_{i=0}^{k-1} : \boldsymbol{y}_1^i=\textit{ESMM}(\boldsymbol{x}, \boldsymbol{W}_1, \boldsymbol{b}_1, \mathcal{R}_i(\boldsymbol{x}))$ \\
            \cmidrule{2-5}
             & $\circled{2}$ & Activation & $\boldsymbol{y}_2=\mathcal{F}(\boldsymbol{y}_1)$ & $\{\boldsymbol{y}_2^i\}_{i=0}^{k-1} : \boldsymbol{y}_2^i=\mathcal{F}(\boldsymbol{y}_1^i)$ \\
            \cmidrule{2-5}
             & $\circled{3}$ & 2nd MLP & $\boldsymbol{y}=\textit{ESMM}(\boldsymbol{y}_2, \boldsymbol{W}_2, \boldsymbol{b}_2, \mathcal{R}(\boldsymbol{x}))$ & $\boldsymbol{y}=\sum_{i=0}^{k-1}\textit{ESMM}(\boldsymbol{y}_2^i, \boldsymbol{W}_2, \boldsymbol{b}_2, \mathcal{R}_i(\boldsymbol{x}))$ \\
            \cmidrule{1-5}
            \multirow{11}{*}{Backward} & \circled{4} & \multirow{4}{*}{2nd MLP} & $\frac{\partial\ell}{\partial \boldsymbol{b}_2}=\textit{ESS}(\frac{\partial\ell}{\partial\boldsymbol{y}}, \mathcal{R}(\boldsymbol{x}))$ & $\frac{\partial\ell}{\partial \boldsymbol{b}_2}=\sum_{i=0}^{k-1}\textit{ESS}(\frac{\partial\ell}{\partial\boldsymbol{y}}, \mathcal{R}_i(\boldsymbol{x}))$ \\
            \cmidrule{2-2}\cmidrule{4-5}
             & \circled{5} &  & $\frac{\partial\ell}{\partial\boldsymbol{W}_1}=\textit{ESTMM}(\boldsymbol{x}, \frac{\partial\ell}{\partial\boldsymbol{y}_1}, \mathcal{R}(\boldsymbol{x}))$ & $\frac{\partial\ell}{\partial\boldsymbol{W}_1}=\sum_{i=0}^{k-1}\textit{ESTMM}(\boldsymbol{x}, \frac{\partial\ell}{\partial\boldsymbol{y}_1^i}, \mathcal{R}_i(\boldsymbol{x}))$ \\
            \cmidrule{2-2}\cmidrule{4-5}
             & \circled{6} &  & $\frac{\partial\ell}{\partial\boldsymbol{y}_2}=\textit{ESMM}(\frac{\partial\ell}{\partial\boldsymbol{y}}, \boldsymbol{W}_2^T, \textit{null}, \mathcal{R}(\boldsymbol{x}))$ & $\Big\{\frac{\partial\ell}{\partial\boldsymbol{y}_2^i}\Big\}_{i=0}^{k-1} : \frac{\partial\ell}{\partial\boldsymbol{y}_2^i}=\textit{ESMM}(\frac{\partial\ell}{\partial\boldsymbol{y}}, \boldsymbol{W}_2^T, \textit{null}, \mathcal{R}_i(\boldsymbol{x}))$ \\
            \cmidrule{2-5}
             & \circled{7} & Activation & $\frac{\partial\ell}{\partial \boldsymbol{y}_1}=\frac{\partial\ell}{\partial\boldsymbol{y}_2}\odot\mathcal{F}^{\prime}(\boldsymbol{y}_1)$ & $\Big\{\frac{\partial\ell}{\partial\boldsymbol{y}_1^i}\Big\}_{i=0}^{k-1} : \frac{\partial\ell}{\partial\boldsymbol{y}_1^i}=\frac{\partial\ell}{\partial\boldsymbol{y}_2^i}\odot\mathcal{F}^{\prime}(\boldsymbol{y}_1^i)$ \\
            \cmidrule{2-5}
             & \circled{8} & \multirow{4}{*}{1st MLP} & $\frac{\partial\ell}{\boldsymbol{b}_1}=\textit{ESS}(\frac{\partial\ell}{\partial\boldsymbol{y}_1}, \mathcal{R}(\boldsymbol{x}))$ & $\frac{\partial\ell}{\boldsymbol{b}_1}=\sum_{i=0}^{k-1}\textit{ESS}(\frac{\partial\ell}{\partial\boldsymbol{y}_1^i}, \mathcal{R}_i(\boldsymbol{x}))$ \\
            \cmidrule{2-2}\cmidrule{4-5}
             & \circled{9} &  & $\frac{\partial\ell}{\partial\boldsymbol{W}_1}=\textit{ESTMM}(\boldsymbol{x}, \frac{\partial\ell}{\partial\boldsymbol{y}_1}, \mathcal{R}(\boldsymbol{x}))$ & $\frac{\partial\ell}{\partial\boldsymbol{W}_1}=\sum_{i=0}^{k-1}\textit{ESTMM}(\boldsymbol{x}, \frac{\partial\ell}{\partial\boldsymbol{y}_1^i}, \mathcal{R}_i(\boldsymbol{x}))$ \\
            \cmidrule{2-2}\cmidrule{4-5}
             & \circled{10} &  & $\frac{\partial\ell}{\partial\boldsymbol{x}}=\textit{ESMM}(\frac{\partial\ell}{\partial\boldsymbol{y}_1}, \boldsymbol{W}_1^T, \textit{null}, \mathcal{R}(\boldsymbol{x}))$ & $\frac{\partial\ell}{\partial\boldsymbol{x}}=\sum_{i=0}^{k-1}\textit{ESMM}(\frac{\partial\ell}{\partial\boldsymbol{y}_1^i}, \boldsymbol{W}_1^T, \textit{null}, \mathcal{R}_i(\boldsymbol{x}))$ \\
            \bottomrule
        \end{tabular}}
    \label{tab:top-k}
\end{table*}

\begin{table*}[ht]
    \centering
    \caption{\textbf{Shape of the thread block and thread grid for expert-specific operators in one MoE layer.} We take top-$1$ routing as an example, where $N^{\prime}$ denotes the length of the re-index vector, which is slightly larger than $N$ and is divisible by $\texttt{BLK}$. Thread blocks are all defined with the same shape to facilitate the fused kernel.}
    \renewcommand\arraystretch{1.2}{
    \small
    \begin{tabular}{l|l|c|c|c|c|c|c}
        \toprule 
        & \text{Operator} & \multicolumn{2}{c|}{Input} & \multicolumn{2}{c|}{Output} & Thread Block & Thread Grid \\
        \midrule
        \multirow{3}*{\rotatebox{90}{\text{forward \; }}} & \multirow{4}*{\textit{ESMM}} & $\boldsymbol{x}$ & $(N, D_1)$ & \multirow{4}*{$\boldsymbol{y}$} & \multirow{4}*{$(N, D_2)$} & \multirow{4}*{\makecell{(\texttt{WARP,}\\ \texttt{TIMES})}} & \multirow{4}*{\makecell{\big($\lceil N^{\prime} / \texttt{BLK}\rceil$, \\ $\lceil D_2/(\texttt{TIMES}\cdot\texttt{BLK})\rceil$\big)}} \\
        \cmidrule{3-4}
        &  & $\boldsymbol{w}$ & $(E, D_1, D_2)$ &  &  &  & \\
        \cmidrule{3-4}
        &  & $\boldsymbol{b}$ & $(E, D_2)$ &  &  &  & \\
        \midrule
        \multirow{9}*{\rotatebox{90}{\text{backward\quad}}} & \multirow{2}*{\textit{ESMM}} & $\boldsymbol{x}$ & $(N, D_2)$ & \multirow{2}*{$\boldsymbol{y}$} & \multirow{2}*{$(N, D_1)$} & \multirow{2}*{\makecell{(\texttt{WARP,}\\ \texttt{TIMES})}} & \multirow{2}*{\makecell{\big($\lceil N^{\prime} / \texttt{BLK}\rceil$, \\ $\lceil D_1/(\texttt{TIMES}\cdot\texttt{BLK})\rceil$\big)}} \\
        \cmidrule{3-4} 
        &  & $\boldsymbol{w}$ & $(E, D_2, D_1)$ &  &  &  & \\
        \cmidrule{2-8}
        & \textit{ESS} & $\boldsymbol{x}$ & $(N, D_2)$ & $\boldsymbol{y}$ & $(E, D_2)$ & \makecell{(\texttt{WARP,}\\ \texttt{TIMES})} & \big($E$, $\lceil D_2/(\texttt{TIMES}\cdot\texttt{BLK})\rceil$\big) \\ 
        \cmidrule{2-8}
        & \multirow{2}*{\textit{ESTMM}} & $\boldsymbol{x}_1$ & $(N, D_1)$ & \multirow{2}*{$\boldsymbol{w}$} & \multirow{2}*{$(E, D_1, D_2)$} & \multirow{2}*{\makecell{(\texttt{WARP,}\\ \texttt{TIMES})}} & \multirow{2}*{\makecell{\big($\lceil D_2/\texttt{BLK}\rceil$, \\ $\lceil D_1/(\texttt{TIMES}\cdot\texttt{BLK})\rceil$, $E$\big)}} \\
        \cmidrule{3-4}
        &  & $\boldsymbol{x}_2$ & $(N, D_2)$ &  &  &  & \\
        \cmidrule{2-8}
        & \multirow{4}*{\textit{ESFK}} & $\boldsymbol{x}_1$ & $(N, D_1)$ & $\boldsymbol{y}_1$ & $(N, D_1)$ & \multirow{4}*{\makecell{(\texttt{WARP,}\\ \texttt{TIMES})}} & \multirow{4}*{\makecell{\big($E$, $\lceil D_1/(\texttt{TIMES}\cdot\texttt{BLK})\rceil$, \\ $\lceil N^{\prime} / \texttt{BLK}\rceil + \lceil D_2/\texttt{BLK}\rceil +$ \\ $\lceil D_2/(\texttt{TIMES}\cdot\texttt{BLK})\rceil$\big)}} \\ 
        \cmidrule{3-6}
        &  & $\boldsymbol{x}_2$ & $(N, D_2)$ & $\boldsymbol{y}_2$ & $(E, D_2)$ &  & \\ 
        \cmidrule{3-6}
        &  & $\boldsymbol{w}_1$ & $(E, D_2, D_1)$ & $\boldsymbol{w}_2$ & $(E, D_1, D_2)$ &  & \\ 
        \bottomrule
    \end{tabular}}
    \label{tab:threads}
    \vspace{-0.5cm}
\end{table*}

\section{Experimental Details}\label{appendix:details}

We provide the details of our CUDA program via enumerating the shape of the thread block and thread grid for expert-specific operators in a single MoE layer in Table~\ref{tab:threads}.

We also provide the PyTorch-style pseudocode for the proxy task we used to examine the computing capacity of the heterogeneous devices, as shown in Algorithm~\ref{alg:test}. We adopt a \texttt{for} loop composed of large matrix multiplications with the same scale as the test program.

\begin{algorithm}[H]
\caption{\textbf{PyTorch-style pseudocode of MoE pipeline.}}
\label{alg:test}
\definecolor{codeblue}{rgb}{0.0,0.0,0.8}
\definecolor{codegreen}{rgb}{0.2,0.6,0.2}
\lstset{
  backgroundcolor=\color{white},
  basicstyle=\fontsize{7.2pt}{7.2pt}\ttfamily\selectfont,
  columns=fullflexible,
  breaklines=true,
  captionpos=b,
  numbers=left,
  numbersep=2pt,
  numberstyle=\tt,
  commentstyle=\fontsize{7.2pt}{7.2pt}\color{codegreen},
  keywordstyle=\fontsize{7.2pt}{7.2pt}\bfseries\color{codeblue},
}
\vspace{-0.2cm}
\begin{lstlisting}[language=python]
    import torch
    import time
    
    device = 'cuda'
    size = 2048
    times = 1024
    
    start_time = time.time()
    for j in range(times):
        mat1 = torch.randn(size, size, device=device)
        mat2 = torch.randn(size, size, device=device)
        y = torch.matmul(mat1, mat2)
    end_time = time.time()
    
    print(end_time - start_time)
\end{lstlisting}
\vspace{-0.2cm}
\end{algorithm}

\section{Experimental Results}\label{appendix:experiments}

We provide the exact values for both memory footprint analysis and average latency analysis in Table~\ref{tab:memory} and \ref{tab:latency}, respectively. Specifically, for latency analysis, we provide the average value for each case with 0.5k, 1k, 1.5k and 2k total steps.


\begin{table*}[ht]
    \centering
    \caption{\textbf{Memory analysis with Tutel, MegaBlocks and \texttt{HEXA-MoE} on Swin-Transformer-MoE benchmark (Base and Small).} Experiments are conducted on 2 homogeneous GPUs with automatic mixed precision in PyTorch and batch size 40 for all the experiments. We set the number of global experts to 8, and record the average GPU memory footprint (GB) on each device.}
    \renewcommand\arraystretch{1.2}{
    \scalebox{0.9}{
    \begin{tabular}{l|c|c|c|c|c|c|c|c|c}
        \toprule 
         & \text{Method} & top-1 & top-2 & top-3 & top-4 & top-5 & top-6 & top-7 & top-8 \\
        \midrule
        \multirow{5}*{\rotatebox{90}{\text{Base}}} & \text{Tutel} & 12.7 & 13.9 & 15.3 & 16.0 & 17.4 & 19.0 & 20.3 & 21.8 \\
        \cmidrule{2-10}
         & \text{MegaBlocks (MoE)} & 13.1 & 13.8 & 14.8 & 15.4 & 16.0 & 16.9 & 17.8 & 18.7 \\
        \cmidrule{2-10}
         & \text{MegaBlocks (dMoE)} & 12.9 & 13.7 & 14.9 & 15.6 & 16.7 & 17.8 & 18.6 & 19.7 \\
        \cmidrule{2-10}
         & \text{Ours (data-centric)} & 10.9 & 11.2 & 11.3 & 11.7 & 12.0 & 12.0 & 12.3 & 12.4 \\
        \cmidrule{2-10}
         & \text{Ours (model-centric)} & 10.0 & 10.2 & 10.3 & 10.6 & 10.5 & 10.7 & 10.9 & 11.4 \\
        \midrule
        \multirow{5}*{\rotatebox{90}{\text{Small}}} & \text{Tutel} & 9.0 & 10.0 & 11.0 & 11.6 & 12.7 & 13.8 & 15.0 & 16.0 \\
        \cmidrule{2-10}
         & \text{MegaBlocks (MoE)} & 9.2 & 9.8 & 10.2 & 10.8 & 11.2 & 11.8 & 12.5 & 13.0 \\
        \cmidrule{2-10}
         & \text{MegaBlocks (dMoE)} & 9.0 & 9.7 & 10.4 & 11.4 & 12.0 & 12.4 & 13.3 & 13.9 \\
        \cmidrule{2-10}
         & \text{Ours (data-centric)} & 8.1 & 8.3 & 8.2 & 8.5 & 8.6 & 8.7 & 9.0 & 9.2 \\
        \cmidrule{2-10}
         & \text{Ours (model-centric)} & 7.7 & 7.8 & 7.8 & 8.0 & 7.9 & 8.2 & 8.3 & 8.5 \\
        \bottomrule
    \end{tabular}}}
    \label{tab:memory}
    \vspace{-0.1cm}
\end{table*}

\begin{table*}[ht]
    \centering
    \caption{\textbf{Latency analysis for Tutel, MegaBlocks and \texttt{HEXA-MoE} on Swin-Transformer-MoE benchmark with base and small scale.} Experiments are conducted on 4 homogeneous GPUs with 4 experts. We set different batch size (bs) for different models under different routing strategy to maximize the utilization of GPU memory. We record the average latency of one step (s) during training.}
    \renewcommand\arraystretch{1.3}{
    \scalebox{0.9}{
    \begin{tabular}{l|c|c|c|c|c|c|c|c|c|c|c}
        \toprule 
        &  & \text{Method} & 0.5k & 1k & 1.5k & 2k &  & 0.5k & 1k & 1.5k & 2k \\
        \midrule
        \multirow{10}*{\rotatebox{90}{\text{Base}}} & \multirow{5}*{\rotatebox{90}{\text{top-1,bs=110}}} & \text{Tutel} & 2.96 & 2.90 & 2.47 & 2.40 & \multirow{5}*{\rotatebox{90}{\text{top-2,bs=100}}} & 2.14 & 2.59 & 2.73 & 2.66 \\
        \cmidrule{3-7}\cmidrule{9-12}
         & & \text{MegaBlocks (MoE)} & 2.10 & 2.66 & 2.57 & 2.43 &  & 2.40 & 2.58 & 2.51 & 2.49 \\
        \cmidrule{3-7}\cmidrule{9-12}
         & & \text{MegaBlocks (dMoE)} & 2.06 & 2.02 & 2.13 & 2.19 &  & 2.47 & 2.72 & 2.63 & 2.55 \\
        \cmidrule{3-7}\cmidrule{9-12}
         & & \text{Ours (model-centric)} & 1.52 & 1.51 & 1.51 & 1.51 &  & 1.61 & 1.60 & 1.60 & 1.60 \\
        \cmidrule{3-7}\cmidrule{9-12}
         & & \text{Ours (data-centric)} & 1.01 & 0.99 & 0.99 & 0.99 &  & 1.17 & 1.16 & 1.16 & 1.16 \\
        \cmidrule{2-12}
         & \multirow{5}*{\rotatebox{90}{\text{top-3,bs=90 }}} & \text{Tutel} & 2.23 & 2.43 & 2.48 & 2.54 & \multirow{5}*{\rotatebox{90}{\text{top-4,bs=80 }}} & 2.27 & 2.34 & 2.20 & 2.18 \\
        \cmidrule{3-7}\cmidrule{9-12}
         & & \text{MegaBlocks (MoE)} & 2.09 & 2.31 & 2.35 & 2.17 &  & 2.11 & 2.18 & 2.05 & 2.03 \\
        \cmidrule{3-7}\cmidrule{9-12}
         & & \text{MegaBlocks (dMoE)} & 2.55 & 2.53 & 2.44 & 2.41 &  & 2.00 & 2.19 & 2.23 & 2.12 \\
        \cmidrule{3-7}\cmidrule{9-12}
         & & \text{Ours (model-centric)} & 1.70 & 1.69 & 1.69 & 1.69 &  & 1.83 & 1.82 & 1.82 & 1.82 \\
        \cmidrule{3-7}\cmidrule{9-12}
         & & \text{Ours (data-centric)} & 1.32 & 1.31 & 1.31 & 1.30 &  & 1.42 & 1.41 & 1.41 & 1.41 \\
        \midrule
        \multirow{10}*{\rotatebox{90}{\text{Small}}} & \multirow{5}*{\rotatebox{90}{\text{top-1,bs=140 }}} & \text{Tutel} & 3.59 & 3.63 & 3.62 & 3.01 & \multirow{5}*{\rotatebox{90}{\text{top-2,bs=130 }}} & 2.83 & 2.96 & 2.72 & 2.53 \\
        \cmidrule{3-7}\cmidrule{9-12}
         & & \text{MegaBlocks (MoE)} & 3.41 & 3.64 & 3.72 & 3.23 &  & 3.04 & 3.26 & 3.11 & 3.02 \\
        \cmidrule{3-7}\cmidrule{9-12}
         & & \text{MegaBlocks (dMoE)} & 3.51 & 3.58 & 3.76 & 3.47 &  & 2.25 & 3.10 & 3.01 & 2.87 \\
        \cmidrule{3-7}\cmidrule{9-12}
         & & \text{Ours (model-centric)} & 1.34 & 1.34 & 1.33 & 1.33 &  & 1.49 & 1.49 & 1.48 & 1.48 \\
        \cmidrule{3-7}\cmidrule{9-12}
         & & \text{Ours (data-centric)} & 0.69 & 0.68 & 0.68 & 0.68 &  & 0.87 & 0.86 & 0.86 & 0.85 \\
        \cmidrule{2-12}
         & \multirow{5}*{\rotatebox{90}{\text{top-3,bs=120 }}} & \text{Tutel} & 2.23 & 2.92 & 3.17 & 3.28 & \multirow{5}*{\rotatebox{90}{\text{top-4,bs=110 }}} & 3.09 & 2.86 & 3.03 & 3.00 \\
        \cmidrule{3-7}\cmidrule{9-12}
         & & \text{MegaBlocks (MoE)} & 2.63 & 2.79 & 2.92 & 3.07 &  & 3.04 & 3.09 & 3.01 & 2.91 \\
        \cmidrule{3-7}\cmidrule{9-12}
         & & \text{MegaBlocks (dMoE)} & 3.08 & 3.24 & 3.38 & 3.09 &  & 3.03 & 3.19 & 3.18 & 2.91 \\
        \cmidrule{3-7}\cmidrule{9-12}
         & & \text{Ours (model-centric)} & 1.61 & 1.60 & 1.59 & 1.59 &  & 1.67 & 1.67 & 1.68 & 1.68 \\
        \cmidrule{3-7}\cmidrule{9-12}
         & & \text{Ours (data-centric)} & 1.07 & 1.05 & 1.05 & 1.05 &  & 1.09 & 1.08 & 1.07 & 1.07 \\
        \bottomrule
    \end{tabular}}}
    \label{tab:latency}
    \vspace{-0.1cm}
\end{table*}


\end{document}